\documentclass{article}
\usepackage{amsmath,amssymb,amsfonts}
\usepackage{nicefrac}
\usepackage{esdiff}
\usepackage{bm} 
\DeclareMathOperator{\dd}{\mathrm{d}}

\DeclareMathOperator{\Covar}{\mathrm{Covar}}
\DeclareMathOperator{\Var}{\mathrm{Var}}
\DeclareMathOperator{\E}{\mathrm{E}}
\DeclareMathOperator*{\argmax}{arg\,max}

\usepackage{enumitem}
\usepackage{numprint}
\usepackage[ruled,lined]{algorithm2e}
\usepackage{subcaption}
\usepackage{float}
\usepackage{graphicx,psfrag,epsf}
\usepackage{pgfplots}

\usepackage{tikz}
\usetikzlibrary{shapes,arrows,positioning,external}
\usetikzlibrary{calc,patterns,decorations.pathmorphing,decorations.markings}

\usepackage[adobe-utopia]{mathdesign} 
\usepackage{crimson} 
\usepackage[scaled]{helvet} 

\usepackage{lineno}
\usepackage[a4paper,width=150mm,top=25mm,bottom=30mm]{geometry} 
\usepackage[onehalfspacing]{setspace}
\usepackage[protrusion=true,expansion=true]{microtype}
\usepackage[english]{babel} 
\usepackage{csquotes} 
\usepackage{hyperref}

\usepackage[style=authoryear,backend=biber, maxbibnames=10]{biblatex}
\usepackage[outdir=./]{epstopdf}
\begin{document}
\noindent\fbox{\parbox{\textwidth}{%
\textbf{Notice:} This is the Author Accepted Manuscript of an article published by Springer in \textit{Statistical Papers}, Vol.~64, pp.~1209--1231, 2023.
The final authenticated version is available online at: \url{https://doi.org/10.1007/s00362-023-01438-9}.

\medskip\noindent
This version of the article has been accepted for publication, after peer review, but is not the Version of Record and does not reflect post-acceptance improvements, or any corrections. Use of this accepted version is subject to the publisher's Accepted Manuscript terms of use: \url{https://www.springernature.com/gp/open-research/policies/accepted-manuscript-terms}.
}}
\vspace{1cm}
\title{Adaptive and robust experimental design for linear {\color{black}dynamical} models using Kalman filter}
	\author{Arno Strouwen*\vspace{-0.3cm} \\
	{\small arno.strouwen@kuleuven.be}\vspace{-0.3cm} \\ 
	{\small ORCID: 0000-0001-8607-4091}\vspace{-0.3cm} \\
	{\small KU Leuven: Department of Biosystems}\vspace{-0.3cm} \\
	{\small Strouwen Statistics}  \vspace{-0.3cm} \\
	$\mbox{ }$\\
	Bart M. Nicolaï\vspace{-0.3cm} \\
	{\small bart.nicolai@kuleuven.be}\vspace{-0.3cm} \\
	{\small ORCID: 0000-0001-5267-1920}\vspace{-0.3cm} \\
	{\small KU Leuven: Department of Biosystems}\vspace{-0.3cm}\\
	$\mbox{ }$\\
	Peter Goos\vspace{-0.3cm} \\
	{\small peter.goos@kuleuven.be}\vspace{-0.3cm} \\
	{\small ORCID: 0000-0002-3854-6506}\vspace{-0.3cm} \\
	{\small KU Leuven: Department of Biosystems}\vspace{-0.3cm} \\
	{\small University of Antwerp: Department of Engineering Management}}
\maketitle
\begin{abstract}
	\noindent
Current experimental design techniques for {\color{black}dynamical} systems often only incorporate measurement noise, while {\color{black}dynamical} systems also involve process noise. To construct experimental designs we need to quantify their information content. The Fisher information matrix is a popular tool to do so. Calculating the Fisher information matrix for linear {\color{black}dynamical} systems with both process and measurement noise involves estimating the uncertain {\color{black}dynamical} states using a Kalman filter. The Fisher information matrix, however, depends on the true but unknown model parameters. In this paper we combine two methods to solve this issue and develop a robust experimental design methodology. First, Bayesian experimental design averages the Fisher information matrix over a prior distribution of possible model parameter values. Second, adaptive experimental design allows for this information to be updated as measurements are being gathered. This updated information is then used to adapt the remainder of the design.
\end{abstract}
\\
\\
\textbf{Acknowledgment:} 
The authors would like to thank the KU Leuven for financial support (project C16/16/002). Author Arno Strouwen thanks the Fund for Scientific Research, Flanders (FWO), project 1S58717N.
\newpage
\section{Introduction}
Control, optimization and analysis of {\color{black}dynamical} systems are increasingly being performed using parametric models \parencite{findeisen}. High-quality data are needed to precisely identify these models. Optimal input design for {\color{black}dynamical} systems deals with the cost-effective collection of these data \parencite{goodwin}.
\\
\\
Most experimental design literature for precisely estimating model parameters of {\color{black}dynamical} systems focuses on models with only measurement noise \parencite{franceschini}, or on models with only process noise, when dealing with autoregressive models for time-series modeling \parencite{hjalmarsson,pintelon}. Relatively little literature exists about designing informative experiments when both measurement and process noise are present. One approach that does combine process and measurement noise for experimental design is that of \textcite{telen2}. These authors use a heuristic extension of the Fisher information matrix used by \textcite{franceschini} to deal with process noise. Our approach differs as we use the formal definition of the Fisher information matrix, based on the variance of the score, which is the gradient of the log-likelihood function. The main challenge that arises in this approach is that estimating the unknown model parameters also requires the hidden {\color{black}dynamical} states to be estimated.
\\
\\
Estimating such hidden states for continuous-time non-linear stochastic differential equations generally has no analytical solution \parencite{solin}. In this paper, we focus on linear discrete-time {\color{black}dynamical} systems with Gaussian measurement and process noise. For these models analytical results exist. Particularly, the Kalman filter is used to estimate the {\color{black}dynamical} state. The Kalman filter has hardly been used in the context of optimal experiments. {\color{black}\textcite{titterington,sagnol} use} the steady state Kalman filter to construct continuous optimal designs, which are asymptotically optimal when a large amount of data is gathered. This is in contrast to exact designs, which are optimized for a finite number of measurements, and which we use in this paper. Because of our focus on a finite number of measurements, our work also does not rely on the steady state prediction error covariance. \textcite{stojanovic} use a robust Kalman filter to generate optimal inputs for autoregressive models with non-Gaussian noise. Instead of autoregressive models, we work with linear state space models, where the matrices describing such a state space model may depend on model parameters that must be estimated as precisely as possible.
\\
\\
The Fisher information matrix (FIM) is a popular tool to quantify the quality of an experiment, as it is related to the inverse of the covariance matrix of the model parameter estimates {\color{black}\parencite{elfving, fedorov3}}. An informative experiment makes a scalar measure of the FIM as large as possible. The major issue with optimal experimental design is the dependence of the FIM, and thus also the optimal inputs for the experiment, on the true, but unknown, model parameters. This presents us with a circular problem as the experiment is needed to precisely estimate the parameters. Locally optimal design, where inputs are optimized for a single initial guess for the parameters, is the traditional method to deal with this issue \parencite{atkinson}. However, this method can be very sensitive to the single initial guess. Generally, there exist two directions to improve on the locally optimal design method, namely robustifying the experiment against the uncertainty in the model parameters and making the experiment adaptive {\color{black}\parencite{pronzato}}.
\\
\\
Robustifying the experiment can be achieved in various ways. One popular approach is min-max experimental design \parencite{wong,korkel}. In this method, the experiment is optimized under the assumption of a worst case scenario. This means that Fisher information matrices are calculated for all elements of a set of possible parameter values and the quality of the experiment is judged based on the least informative matrix in this set. This guarantees that, regardless the true parameter values, the experiment will always have a minimal information content. Another popular approach is {\color{black}pseudo-Bayesian optimal design, which uses an expected value approach \parencite{ryan,chaloner}}. A prior distribution for the model parameters is then used, and the experiment is designed to perform well on average for this prior.
\\
\\
In the second approach to improve on the locally optimal design method, the measurements obtained during the execution of the experiment are used to improve on the initial guess of the model parameters. This is called adaptive or sequential experimental design. A new locally optimal design is then based on the updated model parameters and this process is repeated.
\\
\\
In our work, we combine the {\color{black}pseudo-}Bayesian robustification approach with adaptive experimental design and construct optimal adaptive {\color{black}pseudo-}Bayesian experiments. After every measurement, we update our knowledge of the unknown parameters. A new {\color{black}pseudo-}Bayesian optimal design for the remainder of the experiment is then constructed based on the updated prior distribution. {\color{black}This scheme is similar to model predictive control, where optimal controls are calculated for a horizon into the future, and recalculated whenever new information becomes available \parencite{rawlings}.} Adaptive designs have an additional benefit for {\color{black}dynamical} systems in the presence of process noise. This is because it is difficult to predict the {\color{black}dynamical} state of such systems far into the future, because of the process noise. This causes these future measurements to be uninformative and contributing little to the Fisher information matrix. When adaptively designing an experiment, the estimate of the {\color{black}dynamical} state based on the already gathered data will also reduce the prediction variance of future observations, meaning these future observations become more informative.
\section{Modeling the Information Content of a {\color{black}Dynamical} Experiment}
\subsection{The Model}
Our goal is to find {\color{black}dynamical} inputs $\bm  u_{1:T}$ which lead to {\color{black}the most }precise estimation of the static model parameters $\bm  \theta$ of a linear time-invariant discrete-time {\color{black}dynamical} system with Gaussian noise,
\begin{equation}
	\begin{aligned}
		\bm x_{k} &= F(\bm \theta)\bm x_{k-1} + B(\bm \theta)\bm u_k + \bm w_k,
		\qquad 0<k\leq T \\ 
		\bm y_k &= H(\bm \theta)\bm x_k + \bm v_k.
	\end{aligned}
	\label{system}
\end{equation}
In these equations, $\bm  y_k$ represents the measured outputs at time-step $k$. We assume that the experiment ends after $T$ time-steps, and thus $k$ can range from $1$ to $T$. The measured output is dependent on the {\color{black}dynamical} states $\bm  x_k$ through the output matrix $H(\bm \theta)$. These states completely determine the stochastic evolution of the system over time. The transition of the states from one time-step to the next is impacted by the state matrix $F(\bm \theta)$, as well as the inputs at that time-step, $\bm  u_k$, through the input matrix $B(\bm \theta)$. All three of these matrices $F(\bm \theta)$, $B(\bm \theta)$ and $H(\bm \theta)$  can depend on the model parameters $\bm \theta$. {\color{black}This model is linear in its dynamics, meaning that given $\bm \theta$, $\bm x_{k+1}$ depends linearly on $\bm x_k$ and $\bm u_k$, and similarly $\bm y_k$  depends linearly on $\bm x_k$. The model, however, is not linear in the statistical sense, i.e. the expected values of the measurements are not a linear transformation of the parameters $\bm \theta$.}
\\
\\
Noise is present in both the measurements and the state transitions. We make the following assumptions about the measurement noise $\bm v_k$ and the uncontrollable and unobserved process noise $\bm w_k$ at time-step $k$:
\begin{equation}
	\begin{aligned}
		& \bm w_k \sim \mathcal{N}(\bm 0,Q(\bm \theta)),\\
		& \bm v_k \sim \mathcal{N}(\bm 0,R\bm (\bm \theta)),\\
		&\Covar(\bm w_k,\bm v_l) = \bm 0,\\
		&\Covar(\bm w_k,\bm w_l) = \Covar(\bm v_k,\bm v_l) = \bm 0, \qquad k\ne l.\\
	\end{aligned}
\end{equation}
We thus assume that $\bm v_k$ and $\bm w_k$ both follow a multivariate normal distribution with zero mean and covariance matrices equal to $Q(\bm \theta)$ and $R(\bm \theta)$, respectively. These covariance matrices may also depend on the unknown static parameters $\bm \theta$. The measurement and process noise are independent of each other, and there is also no correlation over time, neither for measurement nor process noise. 
\\
\\
The initial state of the system is also assumed to be multivariate normally distributed, with mean $\bm m_0$ and covariance matrix $P_0$. This state is independent of all later noise. So,
\begin{equation}
	\begin{aligned}
		& \bm x_0 \sim \mathcal{N}(\bm m_0,P_0),\\
		&\Covar(\bm x_0,\bm v_k) = \Covar(\bm x_0,\bm w_k) = \bm 0.
		\label{initial state}
	\end{aligned}
\end{equation}
{\color{black}The description of the dynamical system in Equation (\ref{system}), is popular in the control theory literature. We can also give a purely statistical description of this system.
\begin{equation}
	\begin{aligned}
		& p(\bm x_k |  \bm \theta, \bm x_{k-1}, \bm u_k) \sim \mathcal{N}(\bm F(\bm \theta)\bm x_{k-1} + B(\bm \theta)\bm u_k,Q(\bm \theta)),\\
		& p(\bm y_k | \bm \theta, \bm x_k) \sim \mathcal{N}(\bm H(\bm \theta)\bm x_k,R(\bm \theta)).
		\label{statistical_system}
	\end{aligned}
\end{equation}
This statistical description will be a more useful representation when we move to parameter estimation and experimental design. It is easy to show by induction that the dynamical system is Markovian in the following two ways:
\begin{equation}
	\begin{aligned}
		& p(\bm x_k | \bm \theta, \bm x_{1:k-1}, \bm y_{1:k-1}, \bm u_{1:k}) = p(\bm x_k | \bm \theta, \bm x_{k-1}, \bm u_k),\\
		& p(\bm y_k | \bm \theta, \bm x_{1:k}, \bm y_{1:k-1}, \bm u_{1:k}) = p(\bm y_k | \bm \theta, \bm x_k).
	\end{aligned}
\end{equation}}
\subsection{Parameter Estimation}
Before presenting our experimental design methodology, we first discuss how to estimate the model parameters $\bm \theta$ of the model in Equation  (\ref{system}). One popular approach for parameter estimation is based on the likelihood of the unknown parameters given the observations $\bm y_{1:T}$. Here, $\bm y_{k:l}$ denotes the measured outputs from time-step $k$ to $l$, with both endpoints included and $k \leq l$. We use a similar notation for other vectors.
The log-likelihood after $k$ observations have been collected can be computed with the recursive factorization
\begin{equation}
\begin{aligned}
	L_k(\bm \theta, \bm u_{1:k}, \bm y_{1:k}) &= \log p(\bm y_{1:k}|\bm \theta, \bm u_{1:k})\\
	&=  \log p(\bm y_k|\bm \theta, \bm u_{1:k},\bm y_{1:k-1}) 
	+ \log p(\bm y_{1:k-1}|\bm \theta, \bm u_{1:k-1}).
\end{aligned}
	\label{factorization}
\end{equation}
The first term in this expression can further be computed as
\begin{equation}
	p(\bm y_k|\bm \theta, \bm u_{1:k},\bm y_{1:k-1})
	= \int p(\bm y_k|\bm x_k, \bm \theta)p(\bm x_k|\bm \theta, \bm u_{1:k},\bm y_{1:k-1})\dd \bm x_k.
	\label{likelihood factor}
\end{equation}
In this equation, the first factor of the integrand $p(\bm y_k|\bm x_k, \bm \theta)$ corresponds to $\mathcal{N}(\bm H(\bm \theta)\bm x_k,R(\bm \theta))$, while the second factor, $p(\bm x_k|\bm \theta, \bm u_{1:k},\bm y_{1:k-1})$, called the state predictive distribution, {\color{black}is the normal distribution:}
\begin{equation}
	p(\bm x_k|\bm \theta, \bm u_{1:k},\bm y_{1:k-1}) = \int p(\bm x_k|\bm x_{k-1}, \bm \theta, \bm u_k)
	p(\bm x_{k-1}|\bm \theta, \bm u_{1:k-1},\bm y_{1:k-1})\dd \bm x_{k-1}.
\end{equation}
In this equation, the first factor of the integrand, $p(\bm x_k|\bm x_{k-1}, \bm \theta, \bm u_k)$, {\color{black}is equal to,} $\mathcal{N}(\bm F(\bm \theta)\bm x_{k-1} + B(\bm \theta)\bm u_k, Q(\bm \theta))$, while the second factor, $p(\bm x_{k-1}|\bm \theta, \bm u_{1:k-1},\bm y_{1:k-1})$, is called the state filtering distribution. This distribution can be computed by Bayes' law,
\begin{equation}
	p(\bm x_k|\bm \theta, \bm u_{1:k},\bm y_{1:k}) = 
	\frac{p(\bm y_k | \bm x_k, \bm \theta)
		p(\bm x_k|\bm \theta, \bm u_{1:k},\bm y_{1:k-1})}
	{p(\bm y_k|\bm \theta, \bm u_{1:k},\bm y_{1:k-1})}.
	\label{filtering distribution}
\end{equation}
The denominator of this fraction can be thought of as a normalization factor, ensuring that the state filtering distribution for time-step $k$ integrates to one. The second factor in the numerator is the state predictive distribution. The state filtering distribution thus depends on the state predictive distribution, which in turn depends on the state predictive distribution at the previous time-step. These two recurring equations are known as the Bayesian filtering equations. If the state filtering distribution at time-step $k-1$ is normally distributed as $p(\bm x_{k-1}|\bm \theta, \bm u_{1:k-1},\bm y_{1:k-1)}) =  \mathcal{N}(\bm m_{k-1},P_{k-1})$, then the state predictive distribution is also normally distributed $p(\bm x_k|\bm \theta, \bm u_{1:k},\bm y_{1:k-1}) = \mathcal{N}(F(\bm \theta)\bm m_{k-1} + B(\bm \theta)\bm u_k,F(\bm \theta)P_{k-1}F'(\bm \theta) + Q(\bm \theta))$. If the state predictive distribution is normally distributed as $p(\bm x_{k}|\bm \theta, \bm u_{1:k},\bm y_{1:k-1}) = \mathcal{N}(\bm m_{k}^-,P_{k}^-)$, then the joint state and measurement prediction distribution is also normally distributed,
\begin{equation}
	p(\bm x_k, \bm y_k|\bm \theta, \bm u_{1:k},\bm y_{1:k-1})  = \mathcal{N} \left (
	\begin{bmatrix}
		\bm m_{k}^-\\
		H(\bm \theta)\bm m_{k}^-
	\end{bmatrix},
	\begin{bmatrix}
		P_{k}^- & P_{k}^-H(\bm \theta)'\\
		H(\bm \theta)P_{k}^- & H(\bm \theta)P_{k}^-H(\bm \theta)' + R
	\end{bmatrix}
	\right ).
\end{equation}
The state filtering distribution is then also normal and can be calculated from the conditional distribution of a partitioned multivariate normal distribution \parencite{von}:
\begin{equation}
	\begin{aligned}
		p(\bm x_k|\bm \theta, \bm u_{1:k},\bm y_{1:k}) &= \mathcal{N}(\bm m_k, P_k),\\
		\bm m_k &= \bm m_{k}^- + P_{k}^-H(\bm \theta)'(H(\bm \theta)P_{k}^-H'(\bm \theta) +R(\bm \theta))^{-1} (\bm y_k - H(\bm \theta)\bm m_{k}^-),\\
		P_k &= P_{k}^- - P_{k}^-H(\bm \theta)'(H(\bm \theta)P_{k}^-H'(\bm \theta) +R(\bm \theta))^{-1}H(\bm \theta)P_{k}^-. 
	\end{aligned}
\end{equation}
The first state predictive distribution $p(\bm x_1|\bm \theta, \bm u_1)$ is normally distributed since the initial state distribution $p(\bm x_0)$ is normally distributed, and thus all subsequent state predictive and filtering distributions are normally distributed as well. The above derivation for state predictive and filtering distributions is equivalent to the Kalman filter, with an explicit dependence on the model parameters $\bm \theta$ \parencite{sarkka}. The recursion can thus also be written as:
\begin{equation}
	\begin{aligned}
		\bm m_{k}^- &= F(\bm \theta)\bm m_{k-1} + B(\bm \theta)\bm u_k,\\
		P_{k}^- &= F(\bm \theta)P_{k}F(\bm \theta)' + Q(\bm \theta),\\
		p(\bm x_k|\bm \theta, \bm u_{1:k},\bm y_{1:k-1})  &=  \mathcal{N}(\bm m_{k}^-,P_{k}^-),\\
		\bm v_k &= \bm y_k - H(\bm \theta)\bm m_{k}^-,\\
		S_k &= H(\bm \theta)P_{k}^-H(\bm \theta)' +R(\bm \theta),\\
		K_k &= P_{k}^-H(\bm \theta)'S_{k}^{-1},\\
		\bm m_k &= \bm m_{k}^- + K_k \bm v_k,\\
		P_k &= P_{k}^- - K_k S_k K',\\
		p(\bm x_k|\bm \theta, \bm u_{1:k},\bm y_{1:k}) &= \mathcal{N}(\bm m_k,P_k).\\
	\end{aligned}
	\label{Kalman}
\end{equation}
In these equations $\bm v_k$, $S_k$ and $K_k$ are called the innovation gain residual, innovation gain covariance and optimal Kalman gain, respectively.
The Kalman filter recurses back to the initial state distribution in Equation (\ref{initial state}). Since we know that $p(\bm y_k|\bm x_k, \bm \theta) = \mathcal{N}(\bm H(\bm \theta)\bm x_k,R(\bm \theta))$ and $p(\bm x_k|\bm \theta, \bm u_{1:k},\bm y_{1:k-1})  =  \mathcal{N}(\bm m_{k}^-,P_{k}^-)$, it is easy to see that $p(\bm y_k|\bm \theta, \bm u_{1:k},\bm y_{1:k-1})=\mathcal{N}( H(\bm \theta)\bm m_{k}^-,H(\bm \theta)P_{k}^-H'(\bm \theta) +R(\bm \theta))$. This leads to the following expression for the log-likelihood of the model parameters:
\begin{equation}
	L_k(\bm \theta, \bm u_{1:k}, \bm y_{1:k}) = L_{k-1}(\bm \theta, \bm u_{1:k-1}, \bm y_{1:k-1})
	-\frac{1}{2}\log\left|2\pi S_k\right| -\frac{1}{2}\bm v_{k}' S_{k}^{-1}\bm v_{k},
	\label{log likelihood}
\end{equation}
where $S_k$ and $\bm v_k$ come from the Kalman filter recursion. The likelihood at a certain model parameter value $\bm \theta$ can thus be updated at every time-step by running a Kalman filter, with that particular value of $\bm \theta$.
\subsection{The Fisher Information Matrix}
\label{secFIM3}
One common approach to quantify the quality of the inputs $\bm u_{1:T}$ for precisely estimating the static parameters $\bm \theta$ is the expected Fisher information matrix (FIM):
\begin{equation}
	\begin{aligned}
		\mathcal{I}(\bm \theta,\bm u_{1:T}) &= 
		E_{\bm y_{1:T}|\bm \theta, \bm u_{1:T}} \left ( \frac{\partial \log p(\bm y_{1:T}|\bm \theta, \bm u_{1:T})}{\partial \bm \theta}
		\frac{\partial \log p(\bm y_{1:T}|\bm \theta, \bm u_{1:T})}{\partial \bm \theta}' \right ),\\
		&= -E_{\bm y_{1:T}|\bm \theta,\bm u_{1:T}}\left(\frac{\partial^2 \log p(\bm y_{1:T}|\bm \theta, \bm u_{1:T})}{\partial \bm \theta \partial \bm \theta'}\right).
	\end{aligned}
	\label{expectedFIM}
\end{equation}
In this equation, $\bm y_{1:T}|\bm \theta, \bm u_{1:T}$ denotes the joint distribution of all the measurements, given the parameters $\bm \theta$ and the inputs $\bm u_{1:T}$. In addition, $\frac{\partial \log p(\bm y_{1:T}|\bm \theta, \bm u_{1:T})}{\partial \bm \theta}$ is the gradient of the log-likelihood, and $\frac{\partial^2 \log p(\bm y_{1:T}|\bm \theta, \bm u_{1:T})}{\partial \bm \theta \partial \bm \theta'}$ is the Hessian matrix of the log-likelihood. Formally, the Cramér-Rao bound states that the inverse of the expected FIM is a lower bound, by Loewner ordering, of the covariance matrix of an unbiased estimator of $\bm \theta$. This lower bound {\color{black}determines a hyperellipsoid in the parameter space}. The directions and lengths of the principal axes {\color{black}of this hyperellipsoid} are determined by the eigenvectors and eigenvalues of the inverse of the expected FIM, respectively. The inputs $\bm u_{1:T}$ should thus be chosen such that this hyperellipsoid is as small as possible. {\color{black}One possible way to make the hyperellipsoid small, is by minimizing its volume, this is discussed in more detail in the next section}.
\\
\\
Calculating the expected FIM for arbitrary non-linear models is often intractable, because generally no analytical results are available for the high-dimensional integral that is involved in this calculation. These integrals are then often numerically approximated using Monte Carlo methods \parencite{ryan}. However, our model in Equation (\ref{system}) only contains linear transformations, and multivariate normal distributions remain normal under such transformations. As a result, the measurements $\bm y_{1:T}|\bm \theta, \bm u_{1:T}$ also follow a multivariate normal distribution. More specifically, the expected value and covariance of the measurements can be calculated by using following recursion relations, as adapted from \textcite{cavanaugh} to allow for models with control inputs $\bm u_{1:T}$:
\begin{equation}
	\begin{aligned}
		E(\bm y_r |\bm \theta, \bm u_{1:T})
		&= H(\bm \theta)E(\bm x_r |\bm \theta, \bm u_{1:T}),\\
		E(\bm x_r |\bm \theta, \bm u_{1:T})
		&= F(\bm \theta)E(\bm x_{r-1} |\bm \theta, \bm u_{1:T}) + B(\bm \theta)\bm u_r,\\
		E(\bm x_0|\bm \theta, \bm u_{1:T})
		&= \bm m_0, \\
		\Var (\bm y_r |\bm \theta, \bm u_{1:T})
		&= H(\bm \theta)\Var (\bm x_r |\bm \theta, \bm u_{1:T})H(\bm \theta)' + R(\bm \theta),\\
		\Covar (\bm y_r |\bm \theta, \bm u_{1:T};\bm y_s |\bm \theta, \bm u_{1:T})
		&= H(\bm \theta)F^{r-s}(\bm \theta)\Var (\bm x_s |\bm \theta, \bm u_{1:T})H(\bm \theta)',
		\qquad \forall r>s,\\
		\Covar (\bm y_r |\bm \theta, \bm u_{1:T};\bm y_s |\bm \theta, \bm u_{1:T})
		&=\Covar (\bm y_s |\bm \theta, \bm u_{1:T};\bm y_r |\bm \theta, \bm u_{1:T})', 
		\qquad \forall r<s,\\
		\Var (\bm x_r |\bm \theta, \bm u_{1:T})
		&= F(\bm \theta)\Var (\bm x_{r-1} |\bm \theta, \bm u_{1:T})F(\bm \theta)' + Q(\bm \theta),\\
		\Var (\bm x_0 |\bm \theta, \bm u_{1:T})
		&= P_0.
		\label{recursion0}
	\end{aligned}
\end{equation}
The names Var and Covar are somewhat arbitrary in these equations. For example, if $\bm y_k$ is bivariate, then the matrix $\Var \bm y_k$ is a two by two matrix. We make the distinction between Var and Covar to stress the correlation of measurements over time.
\\
\\
The expected FIM for multivariate normal data is well known \parencite{fedorov}, its $[i,j]$th element is
\begin{equation}
	\label{FIMnormal}
	\begin{aligned}
		\mathcal{I}(\bm \theta,\bm u_{1:T})[i,j]=& 
		\frac{\partial \bm E (\bm y_{1:T}| \bm \theta,\bm u_{1:T})'}{\partial \bm \theta[i]}
		\Covar(\bm y_{1:T}| \bm \theta,\bm u_{1:T})^{-1}\frac{\partial \bm E (\bm y_{1:T}| \bm \theta,\bm u_{1:T})}{\partial \bm \theta[j]}   \\ 
		&+\frac{1}{2}
		\text{tr} \left (
		\Covar(\bm y_{1:T}| \bm \theta,\bm u_{1:T})^{-1}
		\frac{\partial \Covar(\bm y_{1:T}| \bm \theta,\bm u_{1:T})}{\partial \bm \theta[i]} 
		\Covar(\bm y_{1:T}| \bm \theta,\bm u_{1:T})^{-1} 
		\frac{\partial \Covar(\bm y_{1:T}| \bm \theta,\bm u_{1:T})}{\partial \bm \theta[j]}
		\right ).
	\end{aligned}
\end{equation}
In this equation, square brackets are used to select an element of a vector or matrix. This equation does not only involve the expectation and covariance of all the observations, but also the derivative of these quantities with respect to the parameters $\bm \theta$. Similar recursions as in Equation (\ref{recursion0}) exist for these parameter sensitivities. We do not explicitly state these recursions as we calculate them by applying forward mode automatic differentiation on this recursion; see the numerical details in Section \ref{sec:details}. 
\section{D-optimal Experimental Design}
To find an optimal experimental design, we thus have to optimize the inputs $\bm u_{1:T}$ such that the expected FIM is as large as possible, as this leads to {\color{black}the most }precise model parameter estimates. {\color{black}When estimating multiple parameters}, the expected FIM is not a scalar, and we thus need to define what constitutes a large matrix. Since the definition of the expected FIM involves Loewner ordering, it seems natural to also use this ordering to compare the quality of different inputs. However, \textcite{fedorov} demonstrate why it is impossible to directly use this partial ordering of positive semi-definite matrices, and that instead a scalar function of the expected FIM is needed. A popular choice is the determinant of the expected FIM, $\left| \mathcal{I}(\bm \theta,\bm u_{1:T})\right|$. This criterion is known as D-optimality and it is related to the inverse of the volume of the confidence hyperellipsoid.
\subsection{Locally Optimal Experimental Design}
One of the main difficulties in experimental design is the dependence of the expected FIM in Equation (\ref{expectedFIM}) on the true parameters of the system, which are exactly those parameters that the experiment should inform us about. This leads to a circular problem. The simplest way to deal with this issue is by using a single initial guess $\bm \theta^*$ for the parameters at the start of the experiment, and to compute D-optimal designs as
\begin{equation}
	\argmax_{\bm u_\text{min} \leq \bm u_{1:T} \leq \bm u_\text{max}}  \left | \mathcal{I}(\bm \theta^*,\bm u_{1:T}) \right|.
	\label{local optimality}
\end{equation}
This method is called locally optimal design as the design only performs well if this single initial guess for the model parameters is close to the true value. In Equation (\ref{local optimality}), $\bm u_\text{min}$ and $\bm u_\text{max}$ are the minimal and maximal allowed control values, respectively.
\subsection{Robust Experiments}
To make the design more robust, so that it provides {\color{black}a substantial amount of} information if the initial guess is not very close to the true value of the model parameters, we can replace the single initial guess $\bm \theta^*$ with a prior probability distribution $p(\bm \theta)$, which represents our knowledge of possible values of $\bm \theta$ before the experiment has started. We want the experiment to perform well over the parameter values in the domain of $p(\bm \theta)$, where the most likely parameter values have the largest weight. Averaging the determinant of the expected FIM over this prior distribution and then optimizing this average achieves this:
\begin{equation}
		\argmax_{\bm u_\text{min} \leq \bm u_{1:T} \leq \bm u_\text{max}} \int \left | \mathcal{I}(\bm \theta,\bm u_{1:T}) \right | p(\bm \theta) \dd \bm \theta
		\approx \argmax_{\bm u_\text{min} \leq \bm u_{1:T} \leq \bm u_\text{max}} \frac{1}{N} \sum_{i=1}^N
		\left | \mathcal{I}(\bm \theta^i,\bm u_{1:T}) \right |, \qquad \text{draw } \bm \theta^i \text{ from } p(\bm \theta).
		\label{bayesian optimality}
\end{equation}
So, the expectation is approximated by Monte Carlo integration with $N$ draws from the prior distribution $p(\bm \theta)$, each draw having the same weight, $\frac{1}{N}$.
Besides Monte Carlo integration, other methods to numerically calculate this robust D-criterion exist, such as the sigma-point based method of \textcite{telen}. However, these methods complicate the updating of the weights for adaptive experimental design in the following section, and might not be compatible with the jittering described in the discussion section.
\\
\\
Due to the use of a prior distribution, this technique is also called pseudo-Bayesian optimal design \parencite{ryan}. We want to stress that the optimality criterion is only pseudo-Bayesian, and not fully Bayesian. This is because, while the prior information is used to construct the inputs, it is not directly used to influence the estimation of the parameters. Only the information coming from the measurements acquired from the experiment is incorporated in the information criterion.
\subsection{Adaptive Experiments}
\subsubsection{Concept}
In the local optimal design criterion in Equation (\ref{local optimality}) and the pseudo-Bayesian optimal design criterion in Equation (\ref{bayesian optimality}), prior information was only incorporated at the start of the experiment. But as soon as the experiment has started, knowledge is accumulating. That additional information can be exploited to optimize the remainder of the experiment. To formalize this, we now assume that we have already performed an experiment with inputs $\bm u_{1:k}$ and measured outputs  $\bm y_{1:k}$. These measurements are used to form an updated prior distribution after $k$ measurements, $p(\bm \theta | \bm u_{1:k}, \bm y_{1:k})$, which represents our belief in the possible values of the model parameters $\bm \theta$ given the additional information the first $k$ measurements contain. We use this updated prior to optimize the remaining inputs $\bm u_{k+1:T}$ of the experiment:
\begin{equation*}
	\argmax_{\bm u_\text{min} \leq \bm u_{k+1:T}  \leq \bm u_\text{max}} \int \left | \mathcal{I}(\bm \theta,\bm u_{1:T}) \right | p(\bm \theta |  \bm u_{1:k},\bm y_{1:k}) \dd \bm \theta.
\end{equation*}
\subsubsection{Computational Challenges}
This approach, however, has the drawback that optimizing the remaining $N-k$ input vectors for every time-step is computationally much too expensive, especially at the beginning of the experiment, as the behavior of the system then has to be predicted far into the future, using the recursions in equation (\ref{recursion0}). To reduce the computational burden, we only optimize the expected FIM for the next $e$ measurements:
\begin{equation*}
	\argmax_{\bm u_\text{min} \leq \bm u_{k+1:k+e} \leq \bm u_\text{max}} \int \left | \mathcal{I}(\bm \theta,\bm u_{1:k+e}) \right | p(\bm \theta |  \bm u_{1:k}, \bm y_{1:k}) \dd \bm \theta.
\end{equation*}
While this criterion does not require predicting very far into the future, due to the moving horizon $k+1:k+e$, it is still problematic for online computation, as the time and memory required in calculating the expected FIM increases at every time-step, as the dimension of the covariance matrix of $\bm y_{1:k+e}$ keeps growing \parencite{cavanaugh}. This is because, the expected FIM in Equation (\ref{expectedFIM}) involves an expectation over $\bm y_{1:k+e}| \bm \theta, \bm u_{1:k+e}$, even when the outputs $\bm y_{1:k}$ have already been measured.
\subsubsection{Predictive Control}
A computationally feasible alternative approach uses the observed Fisher information matrix, rather than the expected Fisher information matrix. This observed FIM does not require averaging over all possible measurements. Instead, it uses the actually observed values $\bm y_{1:k}$,
\begin{equation}
	\mathcal{J}(\bm \theta,\bm u_{1:k}, \bm y_{1:k}) = 
	-\frac{\partial^2 \log p(\bm y_{1:k}|\bm \theta, \bm u_{1:k})}{\partial \bm \theta \partial \bm \theta'}.
	\label{observedFIM}
\end{equation}
The observed FIM has been argued to be a superior tool to quantify the variance of the model parameter estimates \parencite{efron}, and has already been used in sequential experimental design by \textcite{lane} to produce more precise parameter estimates than a method purely based on the expected FIM. A straightforward solution to keep the optimization cost constant at every time-step would be to combine both the observed FIM, to quantify the information of the $k$ already performed measurements, and the expected FIM, to quantify the information of the $e$ future observations. {\color{black} More theoretically, this can be justified by looking at what the expected value of the observed FIM would be, when averaged over $e$ (unknown) future observations:
\begin{equation}
	\begin{aligned}
	&\E_{\bm y_{k+1:k+e}|\bm \theta, \bm u_{1:k+e}}\mathcal{J}(\bm \theta,\bm u_{1:k+e}, \bm y_{1:k+e}) 
	= -\E_{\bm y_{k+1:k+e}|\bm \theta, \bm u_{1:k+e}}\frac{\partial^2 \log p(\bm y_{1:k+e}|\bm \theta, \bm u_{1:k+e})}{\partial \bm \theta \partial \bm \theta'}
		\\
   &= -\frac{\partial^2 \log p(\bm y_{1:k}|\bm \theta, \bm u_{1:k})}{\partial \bm \theta \partial \bm \theta'}
   -\E_{\bm y_{k+1:k+e}|\bm \theta, \bm u_{1:k+e}}\frac{\partial^2 \log p(\bm y_{k+1:k+e}|\bm \theta,\bm y_{1:k}, \bm u_{1:k+e})}{\partial \bm \theta \partial \bm \theta'}  \\
   &= \mathcal{J}(\bm \theta,\bm u_{1:k}, \bm y_{1:k}) + \mathcal{I}(\bm \theta,\bm u_{k+1:k+e}),
	\end{aligned}	
\end{equation}
where we have used the likelihood decomposition of Equation (\ref{likelihood factor}).
}
To calculate the expected FIM $\mathcal{I}(\bm \theta,\bm u_{k+1:k+e})$ only an expectation over $\bm y_{k+1:k+e}$ is needed. So,
\begin{equation}
	\label{expectedFIMk}
	\begin{aligned}
		&\mathcal{I}(\bm \theta,\bm u_{k+1:k+e})[i,j]= \\
		&\frac{\partial \bm E (\bm y_{k+1:k+e}| \bm \theta, \bm y_{1:k}, \bm u_{1:k+e})'}{\partial \bm \theta[i]}
		\Covar(\bm y_{k+1:k+e}| \bm \theta, \bm y_{1:k},\bm u_{1:k+e})^{-1}\frac{\partial \bm E (\bm y_{k+1:k+e}| \bm \theta, \bm y_{1:k},\bm u_{1:k+e})}{\partial \bm \theta[j]} \\ & \qquad \qquad + \\
		&\qquad\frac{1}{2}
		\text{tr} \left (
		\Covar(\bm y_{k+1:k+e}| \bm \theta, \bm y_{1:k},\bm u_{1:k+e})^{-1}
		\frac{\partial \Covar(\bm y_{k+1:k+e}| \bm \theta, \bm y_{1:k},\bm u_{1:k+e})}{\partial \bm \theta[i]} \right.\\
		&\qquad \left. \qquad\Covar(\bm y_{k+1:k+e}| \bm \theta, \bm y_{1:k},\bm u_{1:k+e})^{-1} 
		\frac{\partial \Covar(\bm y_{k+1:k+e}| \bm \theta, \bm y_{1:k},\bm u_{1:k+e})}{\partial \bm \theta[j]}
		\right ).
	\end{aligned}
\end{equation}
The recursion formulas in Equation (\ref{recursion0}) must thus also be changed to not recurse all the way back to time point $0$. The recursion instead should end at time $k$, with the state mean $\bm m_k$ and covariance estimate $P_k$ coming from the Kalman filter in Equation (\ref{Kalman}):
\begin{equation}
	\begin{aligned}
		&E(\bm y_r |\bm \theta, \bm u_{1:k+e}, \bm y_{1:k})
		= H(\bm \theta)E(\bm x_r |\bm \theta, \bm u_{1:k+e}, \bm y_{1:k}),\\
		&E(\bm x_r |\bm \theta, \bm u_{1:k+e}, \bm y_{1:k})
		= F(\bm \theta)E(\bm x_{r-1} |\bm \theta, \bm u_{1:k+e}, \bm y_{1:k})
		+ B(\bm \theta)\bm u_r,\\
		&\color{blue}E(\bm x_k|\bm \theta, \bm u_{1:k+e}, \bm y_{1:k})
		= \color{blue} \bm m_k,\\
		&\Var (\bm y_r |\bm \theta, \bm u_{1:k+e}, \bm y_{1:k})
		= H(\bm \theta)\Var (\bm x_r |\bm \theta, \bm u_{1:k+e}, \bm y_{1:k})H(\bm \theta)'
		+R(\bm \theta),\\
		&\Covar (\bm y_r |\bm \theta, \bm u_{1:k+e}, \bm y_{1:k};\bm y_s |\bm \theta, \bm u_{1:k+e}, \bm y_{1:k}) =\\
		& \qquad  H(\bm \theta)F^{r-s}\Var (\bm x_s |\bm \theta, \bm u_{1:k+e}, \bm y_{1:k})H(\bm \theta)',
		\quad \forall r>s,\\
		&\Covar (\bm y_r |\bm \theta, \bm u_{1:k+e}, \bm y_{1:k};\bm y_s |\bm \theta, \bm u_{1:k+e}, \bm y_{1:k})=\\
		&\qquad \Covar (\bm y_s |\bm \theta, \bm u_{1:k+e}, \bm y_{1:k};\bm y_r |\bm \theta, \bm u_{1:k+e}, \bm y_{1:k})' ,
		\quad \forall r<s, \\
		&\Var (\bm x_r |\bm \theta, \bm u_{1:k+e}, \bm y_{1:k})
		= F(\bm \theta)\Var (\bm x_{r-1} |\bm \theta, \bm u_{1:k+e}, \bm y_{1:k})F(\bm \theta)'
		+ Q(\bm \theta),\\
		&\color{blue} \Var (\bm x_k |\bm \theta, \bm u_{1:k+e}, \bm y_{1:k})
		= \color{blue} P_k.
	\end{aligned}
	\label{recursioni}
\end{equation}
This leads to the following optimal design criterion at time-step $k$:
\begin{equation}
	\argmax_{\bm u_\text{min} \leq \bm u_{k+1:k+e} \leq \bm u_\text{max}} \int \left |\mathcal{J}(\bm \theta,\bm u_{1:k}, \bm y_{1:k}) + \mathcal{I}(\bm \theta,\bm u_{k+1:k+e}) \right | p(\bm \theta | \bm y_{1:k}, \bm u_{1:k}) \dd \bm \theta.
	\label{adaptive bayesian optimality}
\end{equation}
The integral in Equation (\ref{adaptive bayesian optimality}) can again be approximated by Monte Carlo integration:
\begin{equation}
	\argmax_{\bm u_\text{min} \leq \bm u_{k+1:k+e} \leq \bm u_\text{max}}
	\sum_{i=1}^N
	\left |\mathcal{J}(\bm \theta^i_k,\bm u_{1:k}, \bm y_{1:k}) + \mathcal{I}(\bm \theta^i_k,\bm u_{k+1:k+e}) \right |w_k^i.
	\label{final criterion}
\end{equation}
In this equation, the weights $w_k^i$ and the model parameters $\bm \theta_k^i$ come from the recursion:
\begin{equation}
	\begin{aligned}
		& w_k^i = \frac{p(\bm y_k|\bm \theta_{k-1}^i, \bm u_{1:k},\bm y_{1:k-1})w_{k-1}^i}{\sum_{j=1}^{N}p(\bm y_k|\bm \theta_{k-1}^j, \bm u_{1:k},\bm y_{1:k-1})w_{k-1}^j},\\
		& \bm \theta_k^i = \bm \theta_{k-1}^i,\\
		& w_0^i = \frac{1}{N},\\
		& \bm \theta_0^i \text{ drawn from } p(\bm \theta).
		\label{weight update}
	\end{aligned}
\end{equation}
These weights are thus updated to give higher importance to model parameters according to their likelihood. {\color{black}The recursion for each $\bm \theta^i$ is a constant sequence. This is because our} adaptive experimental design routine only works when using the same Monte Carlo draws $\bm \theta^i_k$ at each time-step. If different values were used at every time-step the likelihoods would have to be calculated again from the beginning, instead of relying on the recursive Equation (\ref{log likelihood}).
\subsubsection{Final Algorithm}
Putting together all these computations, leads to the following Algorithm \ref{algo} which summarizes all the steps of the algorithm for adaptively generating a robust experiment to estimate the model parameters of a linear {\color{black}dynamical} system in the presence of both process noise and measurement noise.\\
\begin{algorithm}[H]
	\SetAlgoLined
	initialize at step zero\\
	\For{i = 1 through N}
	{
		draw $\theta^i_0$ from $p(\bm \theta)$\\
		set initial state distribution $\bm m_0^i = \bm m_0$ and $P_0^i = P_0$\\
		set initial state distribution sensitivities
		$\frac{\partial \bm m_0^i}{\partial \bm \theta}\Bigr|_{\bm \theta = \bm \theta_0^i} = \frac{\partial^2 \bm m_0^i}{\partial \bm \theta \partial \bm \theta'}\Bigr|_{\bm \theta = \bm \theta_0^i} = \frac{\partial P_0^i}{\partial \bm \theta}\Bigr|_{\bm \theta = \bm \theta_0^i} = \frac{\partial^2 P_0^i}{\partial \bm \theta \partial \bm \theta'}\Bigr|_{\bm \theta = \bm \theta_0^i} = 0$
		\\
		set log-likelihood $L(\bm \theta_0^i) = 0$\\
		set observed FIM $\mathcal{J}(\bm \theta_0^i) = 0$\\
		set weight $w_0^i = \frac{1}{N}$\\
	}
	run the experiment\\
	\For{ each time-step k = 0 through T}
	{
		optimize next $e$ controls $\bm u_{k+1:k+e}$ using Eq (\ref{final criterion}), with expected FIM from Eq (\ref{expectedFIMk}) \\
		use control $\bm u_{k+1}$\\
		perform measurement $\bm y_{k+1}$\\
		\For{ i = 1 through N}
		{	Push forward Kalman filter (using hyper-dual numbers) in Eq (\ref{Kalman}) with $\bm \theta_{k-1}^i$\\
			Use results (and intermediate results) from Kalman filter to:\\
			1) update state distribution mean $\bm m_k^i$ and covariance $P_k^i$\\
			2) update state distribution sensitivities  
			$\frac{\partial \bm m_0^i}{\partial \bm \theta}\Bigr|_{\bm \theta = \bm \theta_{k-1}^i}$, 
			$\frac{\partial^2 \bm m_0^i}{\partial \bm \theta \partial \bm \theta'}\Bigr|_{\bm \theta = \bm \theta_{k-1}^i}$,
			$\frac{\partial P_0^i}{\partial \bm \theta}\Bigr|_{\bm \theta = \bm \theta_{k-1}^i}$ and
			$\frac{\partial^2 P_0^i}{\partial \bm \theta \partial \bm \theta'}\Bigr|_{\bm \theta = \bm \theta_{k-1}^i}$\\
			3) update log-likelihood using Eq (\ref{log likelihood})\\
			4) update observed FIM using Eq (\ref{observedFIM})\\
			5) update weights $w_k^i$using Eq (\ref{weight update})\\
			6) update model parameters $\bm \theta_k^i$ using Eq (\ref{weight update})}
	}
	\caption{Robust and adaptive experimental design algorithm for {\color{black}dynamical} systems in the presence of both process and measurement noise}
	\label{algo}
\end{algorithm}
\section{Numerical details}
\label{sec:details}
The entire Algorithm \ref{algo} was implemented in the Julia programming language \parencite{bezanson}. {\color{black}The $[i,j]$th element of the expected FIM is given in Equation (\ref{expectedFIMk}). However a batch form, where all elements of this matrix are calculated at once, is used in practice, see \textcite{fedorov} for the details. Similarly, the recursion in Equation (\ref{recursioni}) can be efficiently calculated in batch form, we give the equation for the covariance between the $r$'th and $s$'th observation, but in practice a giant covariance matrix between all observations is constructed. This batch form can easily be adapted from the results in \textcite{cavanaugh}.}
\\
\\
For the sensitivities required to calculate the expected FIM, forward mode automatic differentiation is used. In forward mode automatic differentiation, every variable is replaced with a dual number containing both the value of that number and the partial derivatives of that variable with respect to $\bm \theta$. Operators are then overloaded to correctly propagate the partial derivatives \parencite{griewank}. For example, if in the original code there is an expression $c=a*b$, and we know the partial derivatives of $a$ and $b$, these numbers are replaced by the dual numbers $(a,\frac{\partial a}{\partial \bm \theta})$ and $(b,\frac{\partial b}{\partial \bm \theta})$ and multiplication is overloaded as $(a,\frac{\partial a}{\partial \bm \theta})*(b,\frac{\partial b}{\partial \bm \theta})=(a*b,a*\frac{\partial b}{\partial \bm \theta} + b*\frac{\partial a}{\partial \bm \theta}) = (c,\frac{\partial c}{\partial \bm \theta})$. Variables that do not depend on $\bm \theta$ have zero partial derivatives, and the $i$th element of $\bm \theta$ is initialized with one for the $i$th partial derivative and zero for the other partial derivatives. For the observed FIM, second order derivatives are needed. These can also be calculated using forward mode automatic differentiation using hyper-dual numbers, see \textcite{revels} for details how these are implemented in Julia.
\\
\\
Sequential quadratic programming, as implemented in NLopt \parencite{johnson}, is used to solve the optimization problem \parencite{kraft,kraft2} in Equation (\ref{final criterion}). The optimal controls found at the previous time-step, are reused as a hot starting point. A random value between $\bm u_\text{min}$ and $\bm u_\text{max}$ is selected for the controls at the end of the optimization horizon. The optimization algorithm is allowed a maximum of $20$ function evaluations before termination, except for the first time-step where $120$ evaluations are allowed. Gradients of the control objective are again calculated using hyper-dual numbers. 
\section{Case Studies}
\subsection{Mass-Spring-Damper System}
\subsubsection{Problem description}
\begin{figure}[H]
	\centering
	
	\begin{tikzpicture}
		\tikzstyle{spring}=[thick,decorate,decoration={zigzag,pre length=0.3cm,post length=0.3cm,segment length=6}]
		\tikzstyle{damper}=[thick,decoration={markings,  
			mark connection node=dmp,
			mark=at position 0.5 with 
			{
				\node (dmp) [thick,inner sep=0pt,transform shape,rotate=-90,minimum width=15pt,minimum height=3pt,draw=none] {};
				\draw [thick] ($(dmp.north east)+(2pt,0)$) -- (dmp.south east) -- (dmp.south west) -- ($(dmp.north west)+(2pt,0)$);
				\draw [thick] ($(dmp.north)+(0,-5pt)$) -- ($(dmp.north)+(0,5pt)$);
			}
		}, decorate]
		\tikzstyle{ground}=[fill,pattern=north east lines,draw=none,minimum width=0.75cm,minimum height=0.3cm,inner sep=0pt,outer sep=0pt]
		
		\node [style={draw,outer sep=0pt,thick}] (M) [minimum width=1cm, minimum height=2.5cm] {$M$};
		
		\draw [thick] (M.south west) ++ (0.2cm,-0.125cm) circle (0.125cm)  (M.south east) ++ (-0.2cm,-0.125cm) circle (0.125cm);
		
		\node (wall) [ground, rotate=-90, minimum width=3cm,yshift=-3cm] {};
		\node (y) at (M.south) [yshift = -0.25cm] {$y$};
		
		\draw [spring] (wall.170) -- node [above] {$K$} ($(M.north west)!(wall.170)!(M.south west)$);
		\draw [damper] (wall.10) -- node [above=0.25cm] {$C$} ($(M.north west)!(wall.10)!(M.south west)$);
		\draw [-latex,ultra thick] (wall.0) -- (y);
		
		\node (F) at (M.east)[xshift = 2cm] {$F$};
		\draw [-latex,ultra thick] (M.east) -- (F.west);
	\end{tikzpicture} 
	\caption{Schematic representation of mass spring damper system.} 
	\label{fig:msd}
\end{figure}
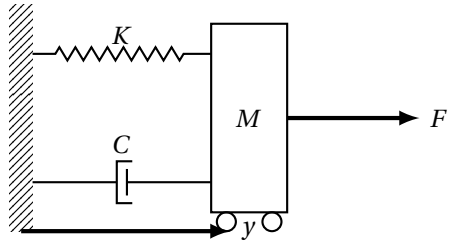

In the first case study, we consider experimental design for the mass-spring-damper system depicted in Figure \ref{fig:msd}. The discreet linear dynamics of this system are:
\begin{equation}
	\begin{aligned}
		&\bm x_k = \begin{bmatrix}
			1 & \Delta t\\
			\frac{-\Delta tK}{M} & \frac{-\Delta tC}{M} + 1
		\end{bmatrix} \bm x_{k-1} +
		\begin{bmatrix} 0 \\ \frac{\Delta t}{M}\end{bmatrix} F_k + 
		\bm w_k,\\
		&\bm w_k \sim  \mathcal{N}\left(\begin{bmatrix}0\\0\end{bmatrix},
		\begin{bmatrix}\frac{q\Delta t^3}{3M^2} & \frac{q\Delta t^2}{2M^2} \\\frac{q\Delta t^2}{2M^2}  & \frac{q \Delta t}{M^2} \end{bmatrix}\right),\\
		&y_k = \begin{bmatrix} 1 & 0 \end{bmatrix} \bm x + v_k,\\
		&v_k \sim \mathcal{N}(0,0.1).
	\end{aligned}
\end{equation}
In these equations, $K$ and $C$ are the spring and damper constant, respectively. These are the two unknown model parameters that must be estimated. Their true values are equal to $1$ and $2$, respectively. The prior distributions we use for them are independent normal distributions centered around $1.4$ and $4$, with variances equal to $0.2$ and $2${\color{black}, respectively}. The parameters $q$ and $M$ are the spectral density of the process noise and mass respectively, which {\color{black}are} known and equal to $0.05$ and $1$. Finally, $\Delta t$ is the time between measurements, equal to $0.1$. The position and velocity are the two states, but only the position is measured, with measurement noise on top of it. The initial state distributions are independent normal distributions with means equal to zero and variances equal to $0.1$. The controllable input at the $k$th time-step is a force $F_k$, which must be optimized such that $K$ and $C$ can be estimated as precisely as possible from the position measurements. The maximum absolute value of the force that can be applied is $1$.
\subsubsection{Optimal Versus Random Design}
\begin{figure}[H]
	\begin{subfigure}[b]{0.90\textwidth}
		\centering
		\includegraphics[width=0.8\textwidth]{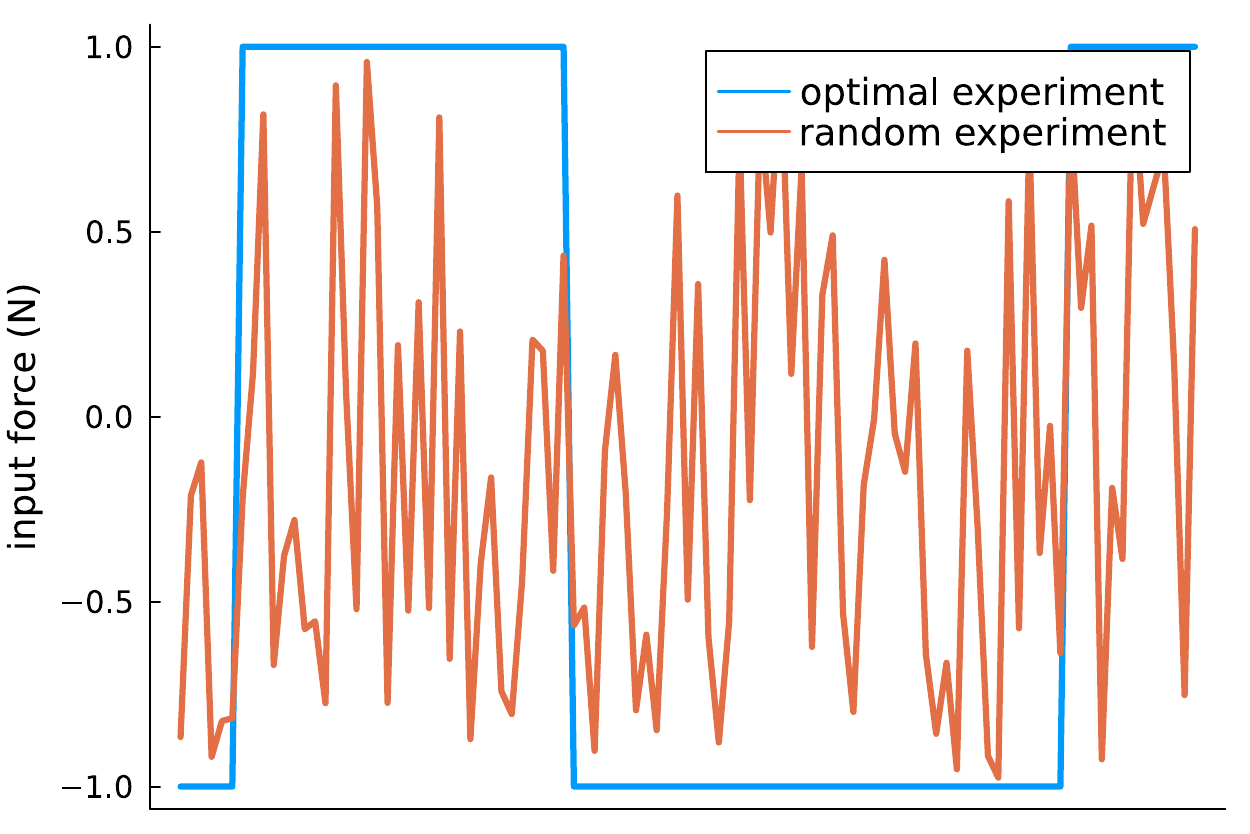}
		\caption{Input.}
		\label{figInput}
	\end{subfigure}\\
	\begin{subfigure}[b]{0.90\textwidth}
		\centering
		\includegraphics[width=0.8\textwidth]{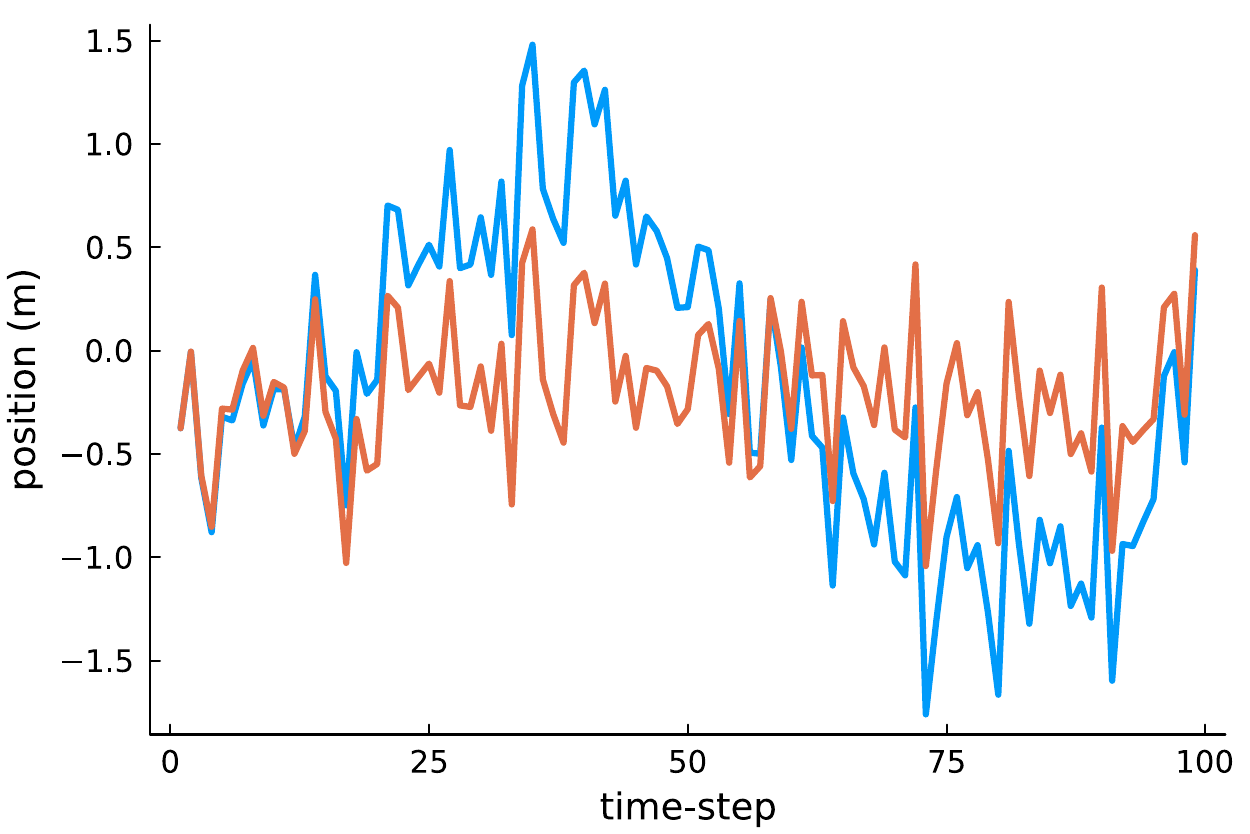}
		\caption{Output.}
		\label{figOutput}
	\end{subfigure}
	\caption{Comparison of optimal inputs compared to random inputs and the corresponding output behaviors.} 
\end{figure}

\begin{figure}[H]
	\centering
	\includegraphics[width=0.8\textwidth]{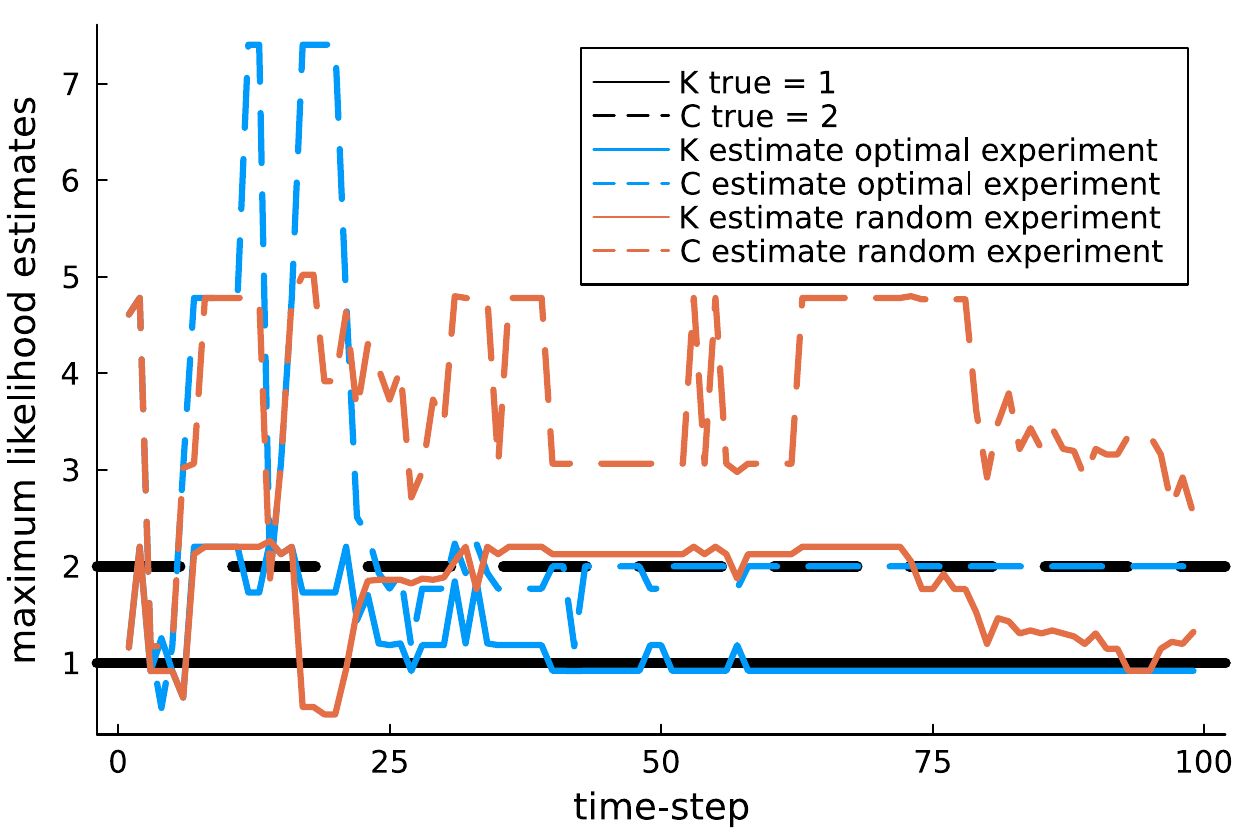}
	\caption{Online maximum likelihood estimates for both the optimal experiment and the random experiment. The optimal experiment converges faster to the true parameters.}
	\label{figOnline}
\end{figure}

\begin{figure}[H]
	
	\begin{subfigure}[b]{0.45\textwidth}
		\includegraphics[width=1.0\textwidth]{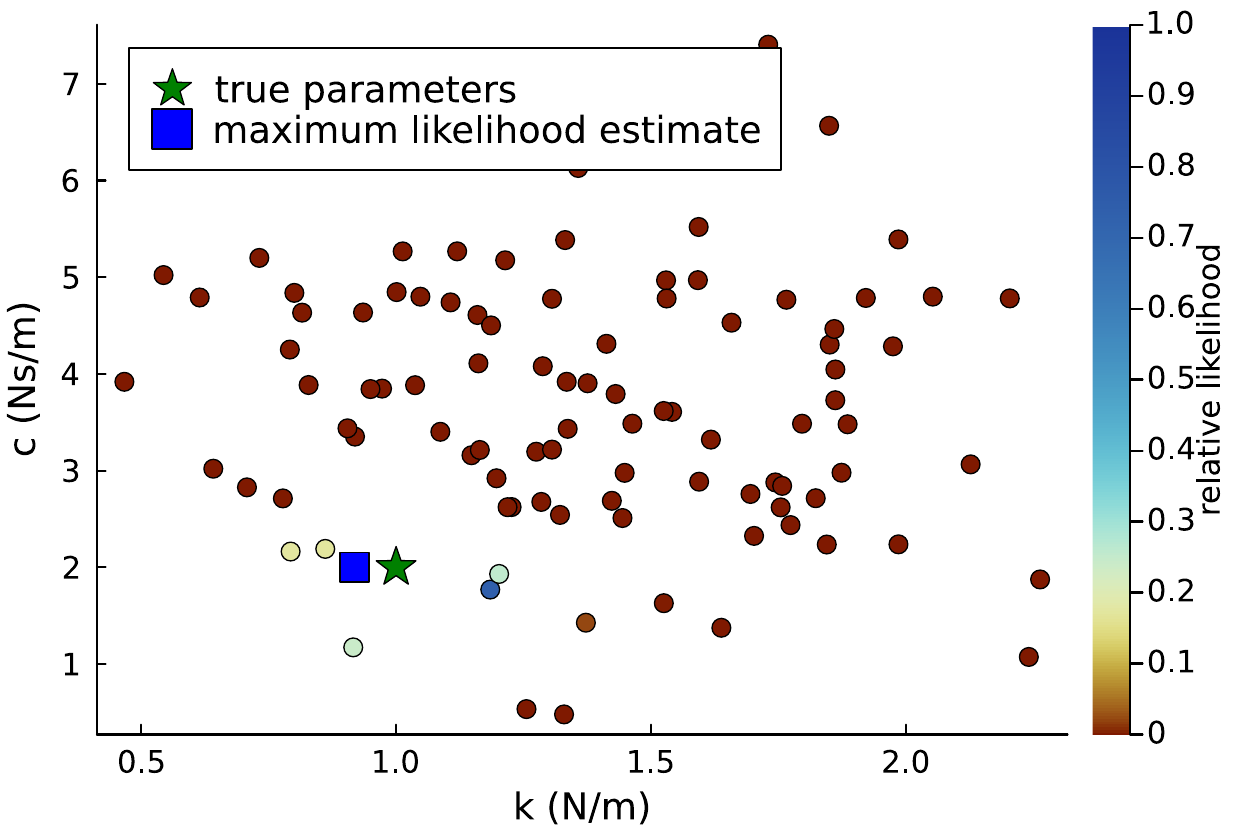}
		\caption{Optimal experiment.}
		\label{figLikelihoodOpt}
	\end{subfigure}
	\begin{subfigure}[b]{0.45\textwidth}
		\centering
		\includegraphics[width=1.0\textwidth]{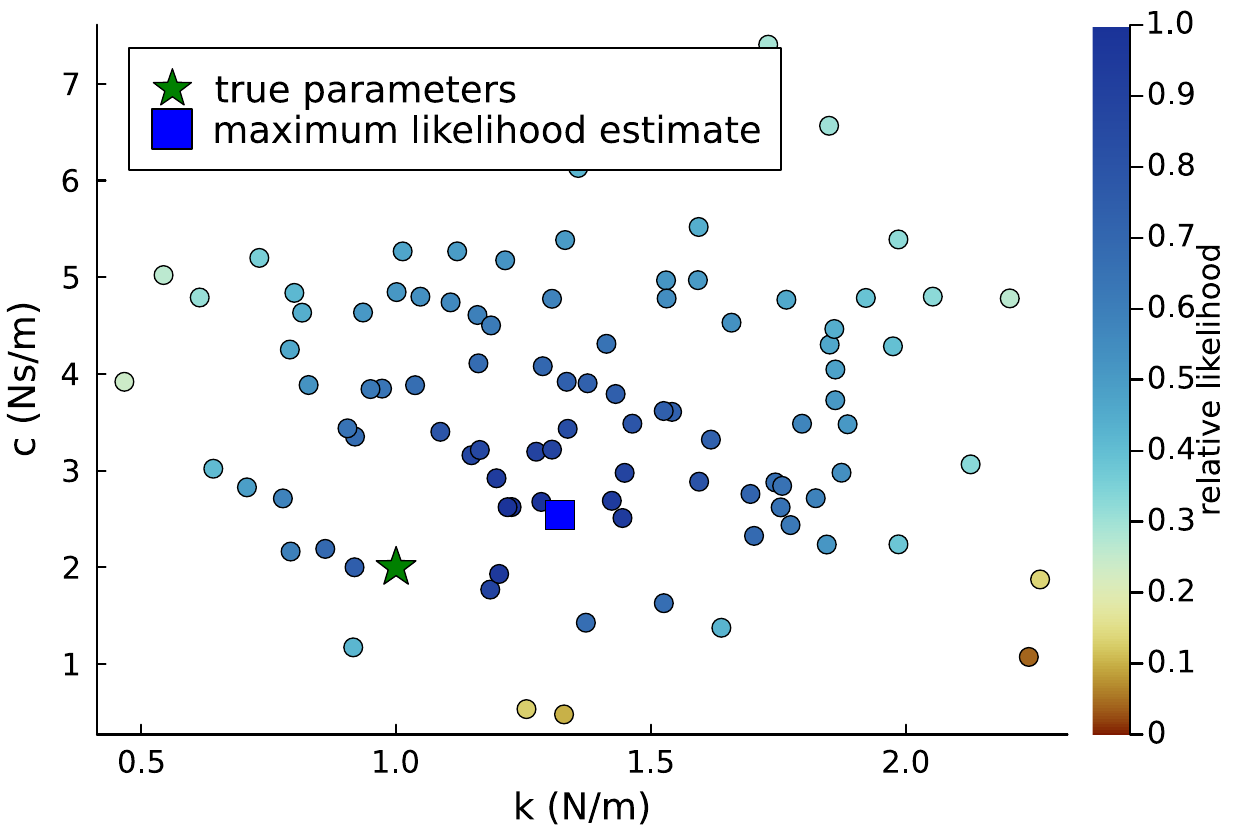}
		\caption{Random experiment.}
		\label{figLikelihoodRand}
	\end{subfigure}
	\caption{{\color{black}Likelihood at the end of the experiment, evaluated at $100$ values of $\bm \theta$, drawn from $p(\bm \theta)$}. The likelihood decreases sharply away from the true parameters for the optimal experiment, unlike in the random experiment. The maximum likelihood estimate of the optimal experiment is much closer to the true value than the random experiment.} 
	\label{figLikelihood}
\end{figure}

We start by comparing the performance of our optimal experimental design strategy to a random input signal. Both experiments last $T=100$ time-steps. The optimal experiment is generated with $N=100$ draws from the prior distribution of the model parameters, and looks $e=3$ steps ahead for optimizing the controls. In Figure \ref{figInput}, the inputs for both experiments are shown, and the corresponding measurements are shown in Figure \ref{figOutput}. An always maximal or minimal control action (bang-bang control) seems to be preferred, since the optimal experimental design switches thrice between the maximum and minimum allowed force. The influence of these optimal controls is clearly visible on the measurements, where the position is clearly lower after a negative force has been applied, and clearly higher after a positive force has been applied. The controls seem to switch from positive to negative and vice versa after the position stagnates around position values of $1$ and $-1$. This is logical since, once the position stagnates, nothing can be learned anymore about the damping constant.
In Figure \ref{figOnline},  we show the evolution of the online maximum likelihood estimates as the experiment progresses. The optimal experiment hovers around the true model parameters after only $50$ time-steps, while the random experiment can not even correctly estimate these parameters after $100$ time-steps. {\color{black}Note that even for the optimal experiment the estimates are not exactly equal to the true values, this is because we can only evaluate the likelihood at the $N$ draws from $p(\bm \theta)$. The estimate can thus at best converge to the draw that was closest to the true values.}
The likelihood at the end of the experiment for the $100$ parameters drawn from the prior distribution for the model parameters is shown in Figure \ref{figLikelihood}. We see that the maximum likelihood estimate for the optimal experiment is one of the closest grid points to the true model parameter values, while this is not the case for the random experiment. Furthermore, for the optimal experiment the relative likelihood of other model parameters compared to the maximum likelihood estimate decreases rapidly when moving away from this estimate. This means that for the optimal experiment only model parameter values close to the true values fit the data well. For the random experiment the likelihood does not decrease rapidly when moving away from the maximum likelihood estimate, which means almost all values fit the data almost equally well and we can not discern the true model parameter values from the data.
\subsubsection{Added Value of the Robustness and Adaptivity}
\begin{figure}[H]
	\begin{subfigure}[b]{0.45\textwidth}
		\includegraphics[width=1.0\textwidth]{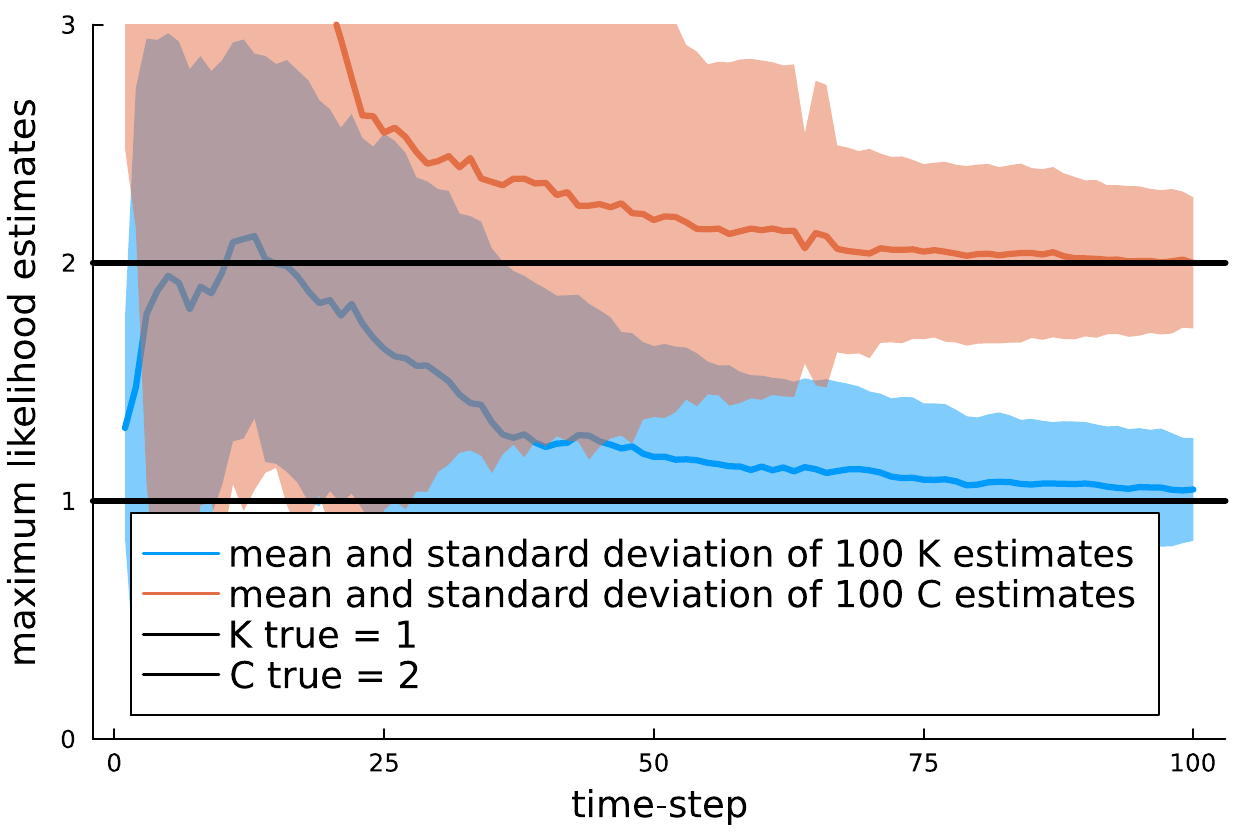}
		\caption{Optimal experiment: $e=3$, $N=100$.}
		\label{figEnsembleOpt}
	\end{subfigure}
	\begin{subfigure}[b]{0.45\textwidth}
		\centering
		\includegraphics[width=1.0\textwidth]{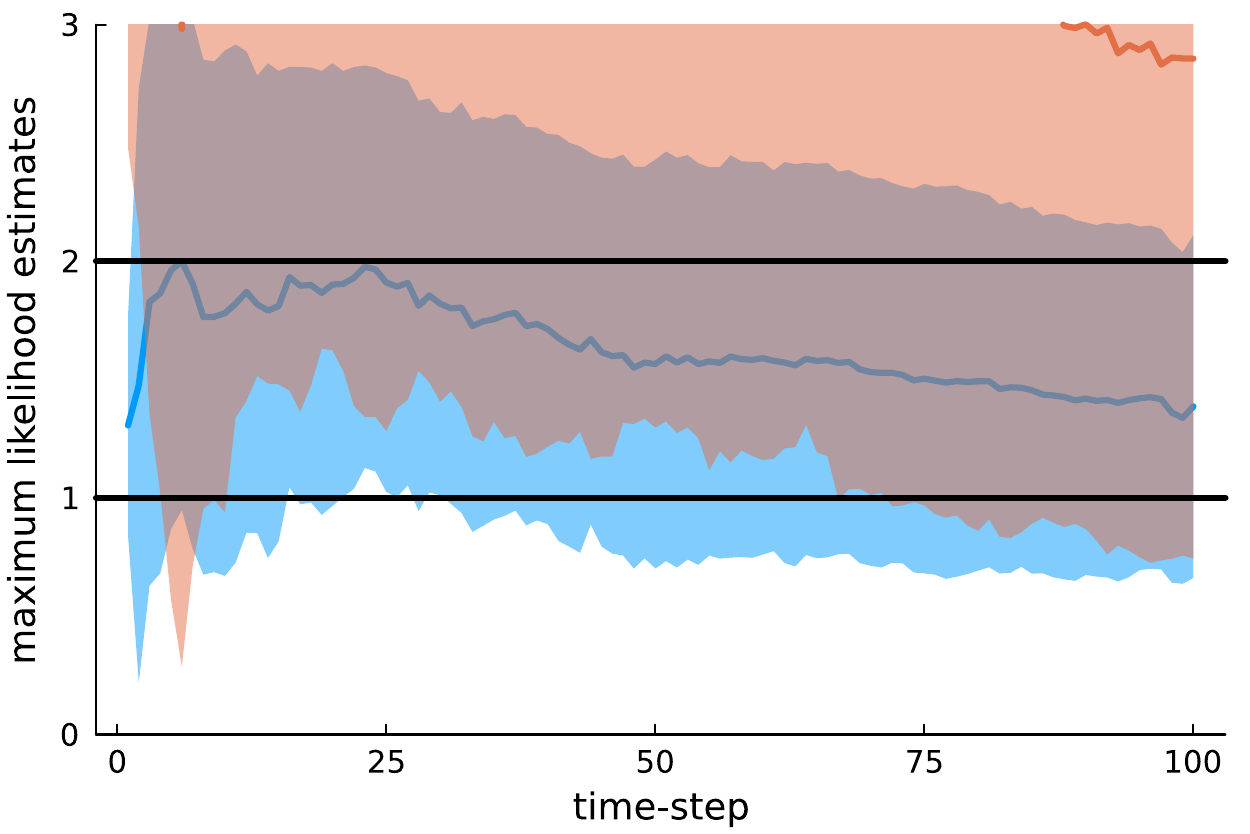}
		\caption{Random experiment.}
		\label{figEnsembleRand}
	\end{subfigure}\\
	\begin{subfigure}[b]{0.45\textwidth}
		\includegraphics[width=1.0\textwidth]{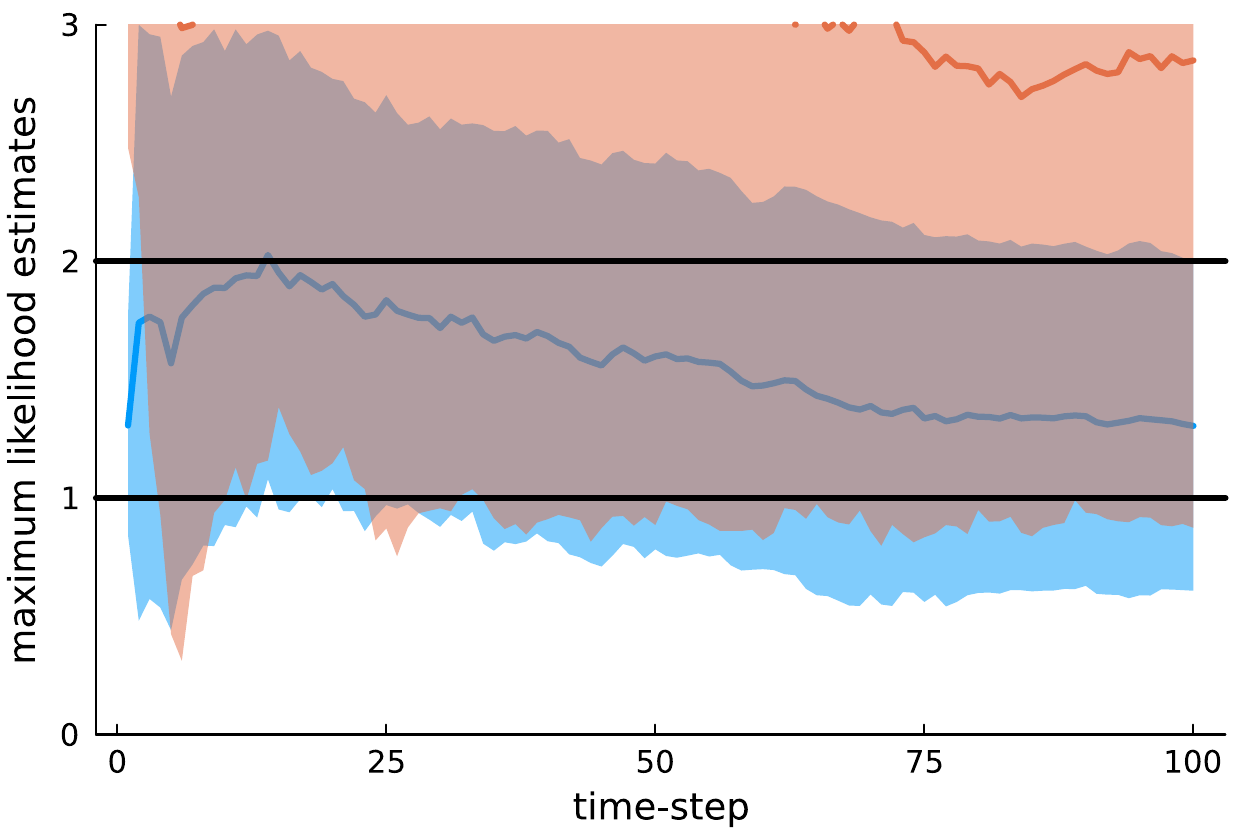}
		\caption{Reduced control horizon: $e=1$, $N=100$.}
		\label{figEnsembleRedHor}
	\end{subfigure}
	\begin{subfigure}[b]{0.45\textwidth}
		\centering
		\includegraphics[width=1.0\textwidth]{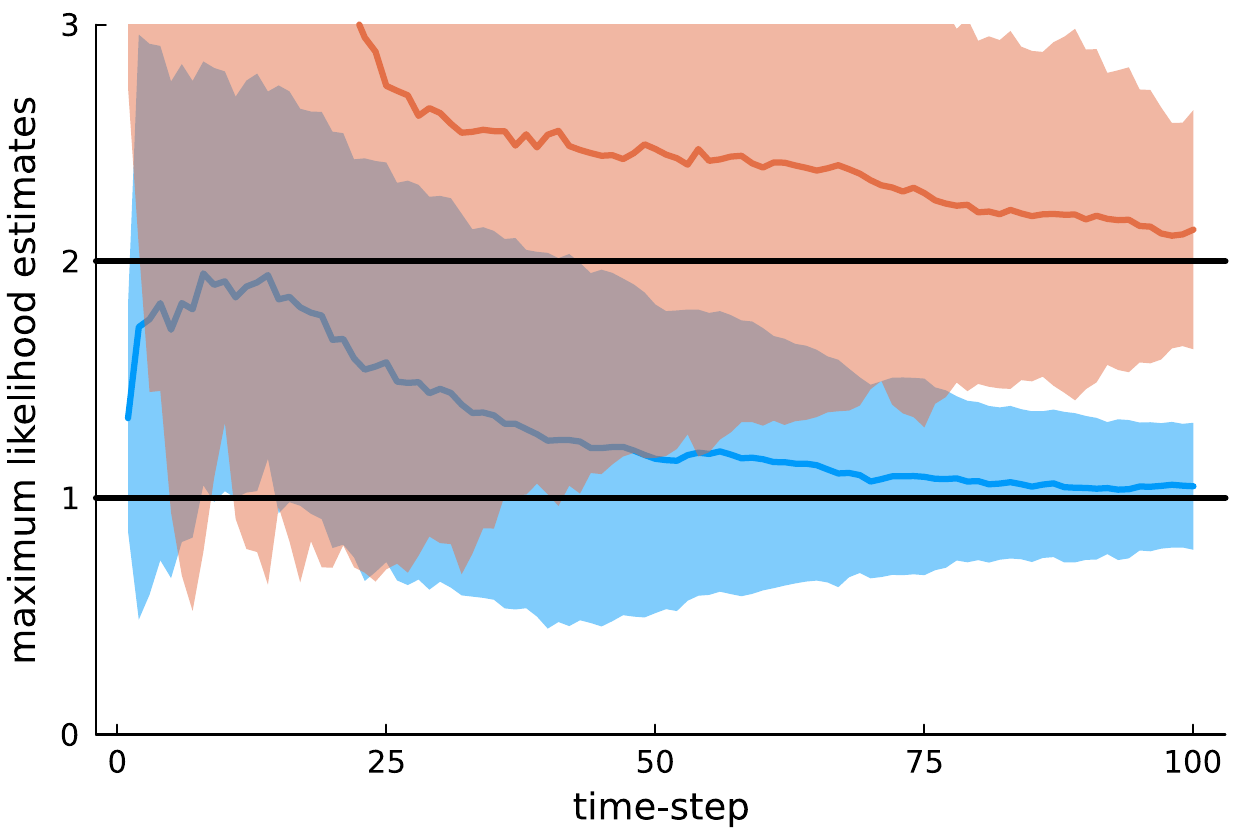}
		\caption{Increased control horizon: $e=6$, $N=100$.}
		\label{figEnsembleIncrHor}
	\end{subfigure}\\
	\begin{subfigure}[b]{0.45\textwidth}
		\includegraphics[width=1.0\textwidth]{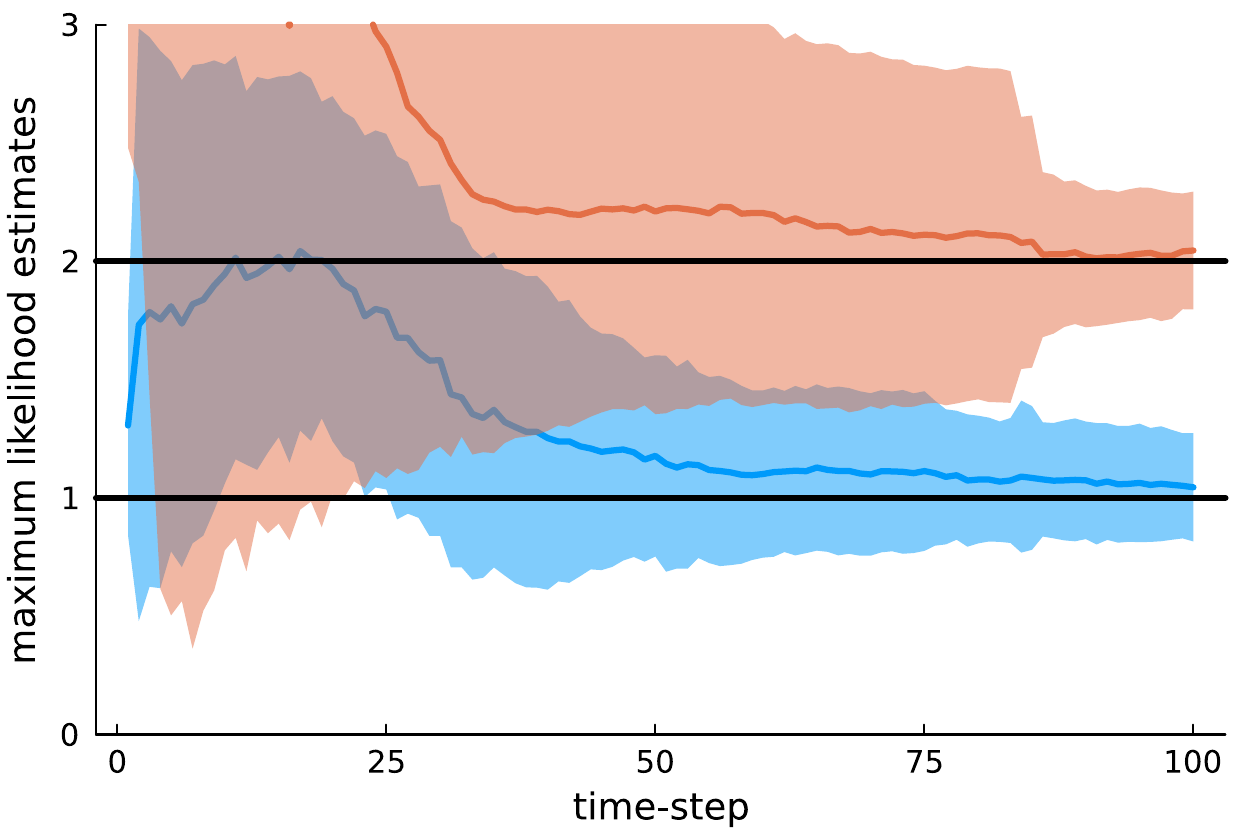}
		\caption{Non-robustness experiment: $e=3$, $N=1$.}
		\label{figEnsembleRedRob}
	\end{subfigure}
	\begin{subfigure}[b]{0.45\textwidth}
		\centering
		\includegraphics[width=1.0\textwidth]{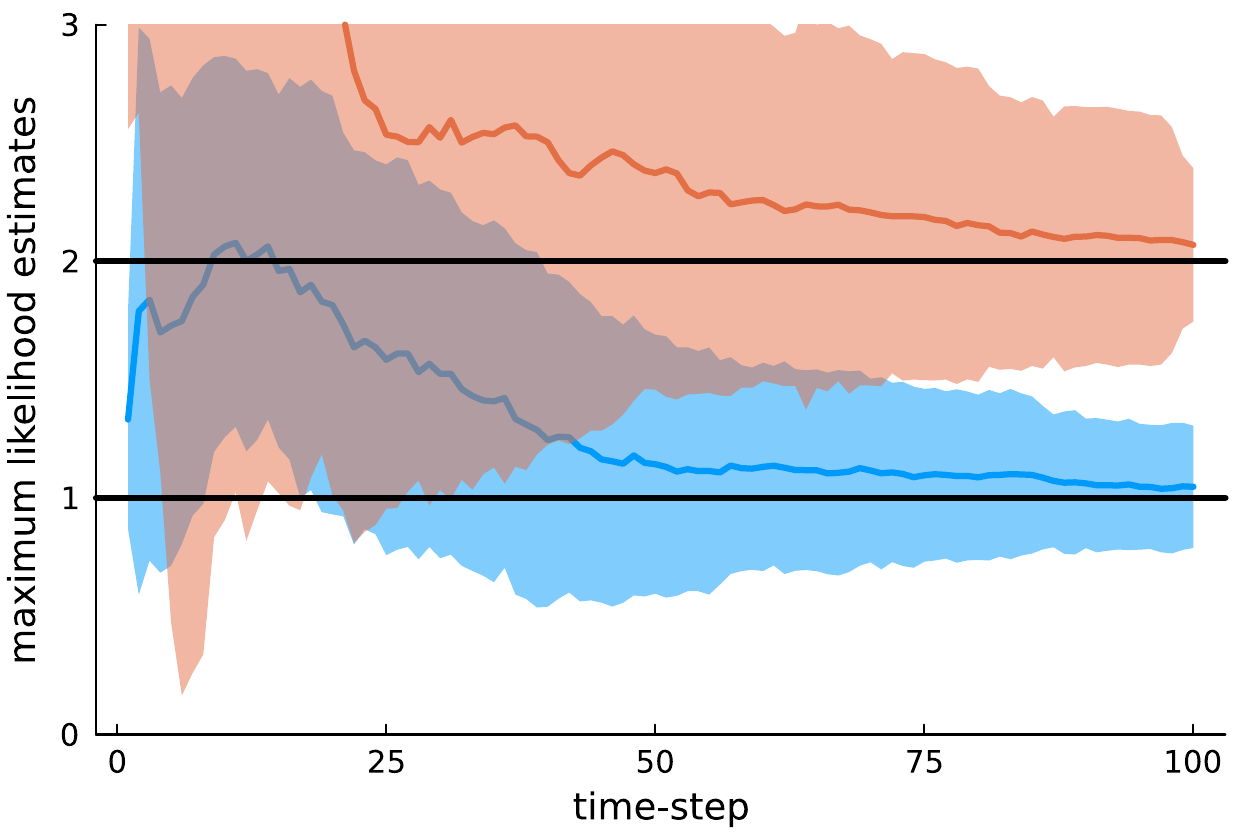}
		\caption{Increased robustness: $e=3$, $N=400$.}
		\label{figEnsembleIncrRob}
	\end{subfigure}\\
	\begin{subfigure}[b]{0.45\textwidth}
		\includegraphics[width=1.0\textwidth]{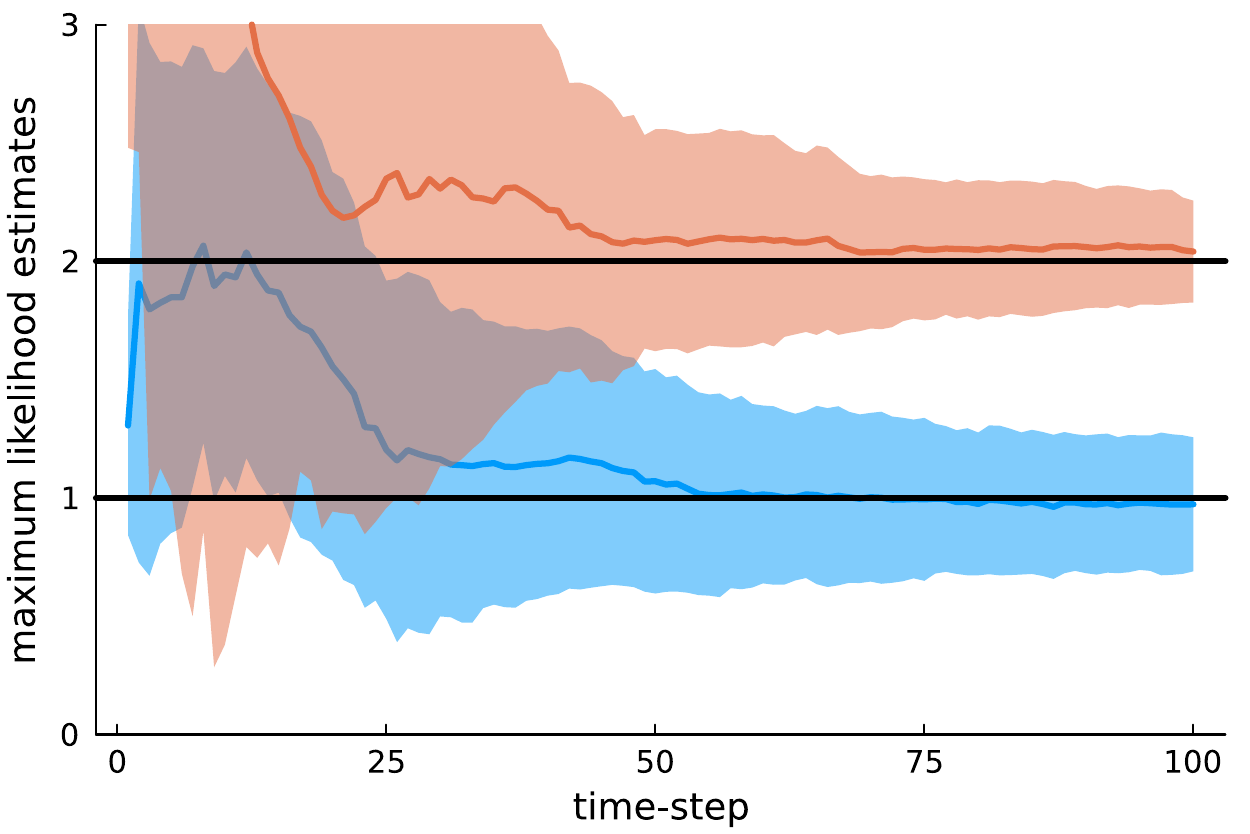}
		\caption{Non-adaptive experiment: $N=100$.}
		\label{figEnsembleNonAdp}
	\end{subfigure}
	\begin{subfigure}[b]{0.45\textwidth}
		\centering
		\includegraphics[width=1.0\textwidth]{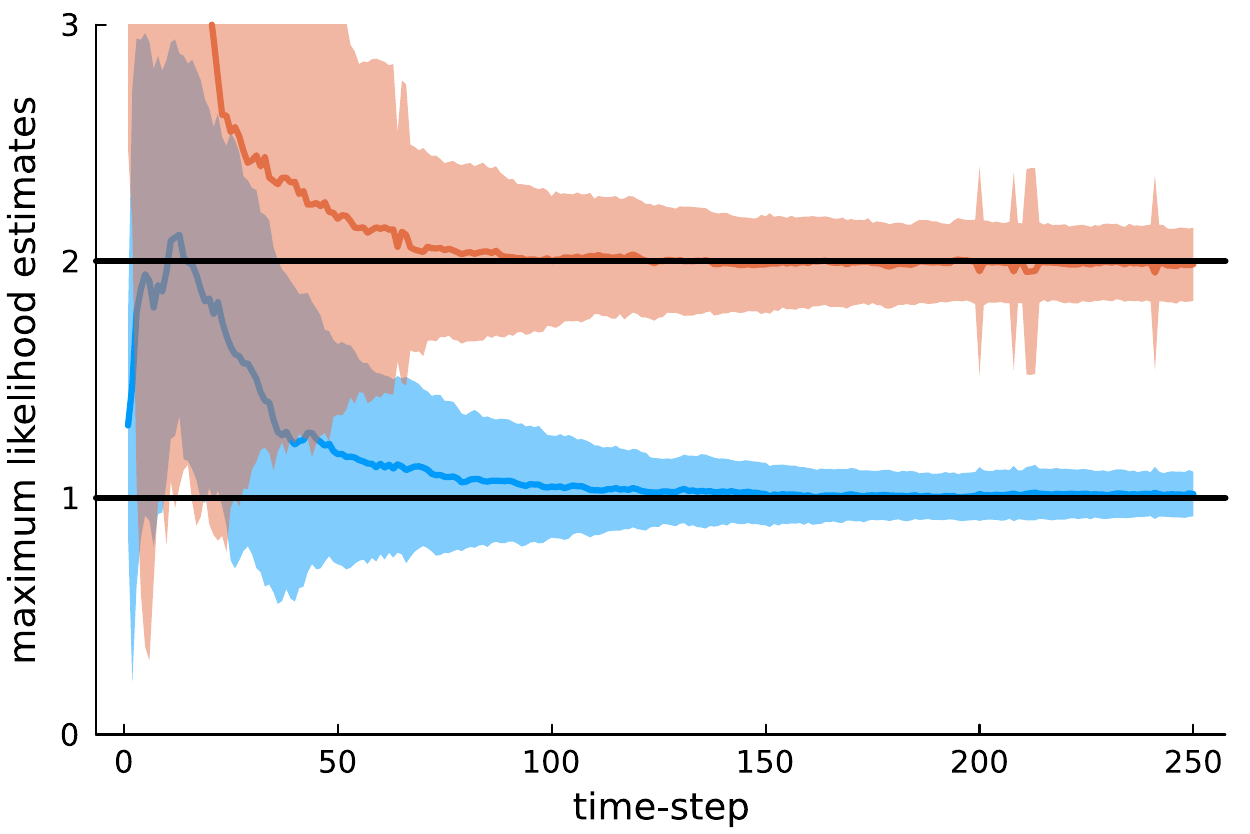}
		\caption{Increased time: $e=3$, $T=100$.}
		\label{figEnsembleIncrTime}
	\end{subfigure}
	\caption{\color{black}$100$ experiments are performed for different combinations of $e$, $N$ and $T$ in Algorithm \ref{algo}. The mean and standard deviation of the online maximum likelihood estimates over these experiments are tracked.} 
	\label{fig:analysis}
\end{figure}

The above discussion already shows the value of experimental design methodology compared to random inputs. We now continue by showing the combined added value of robustness and adaptivity. In Figure \ref{fig:analysis}, we study the behavior of Algorithm \ref{algo} for a variety of combinations of control horizon length $e$, number of model parameters drawn from the prior $N$, and number of time-steps, $T$. For each combination of $e$, $N$ and $T$ the experiment is repeated $100$ times, {\color{black}each with a different realization of the process and measurement noise. For each experiment the maximum likelihood estimate is tracked, and at each time-step the mean and standard deviation of the 100 maximum likelihood estimates are plotted.} Figures \ref{figEnsembleOpt} and \ref{figEnsembleRand} show the same combination as was used before, in {\color{black}Figure \ref{figOnline}}, for the optimal experiment and random experiment, respectively. This allows us to confirm that the optimal experiment performs much better than the random experiment over an ensemble of $100$ experiments, and, thus, that the better estimates of the experimental design methodology were not by chance.
\\
\\
In Figure \ref{figEnsembleRedHor}, the control horizon length, $e$, is reduced from $3$ to $1$. This causes this experiment to perform almost as bad as the random experiment. Increasing the control horizon length to $6$, however, does not greatly increase the performance of the experiment, as shown in Figure \ref{figEnsembleIncrHor}. {\color{black}In fact, the results for $e=6$ look slightly worse than $e=3$. We hypothesize this is because the solver makes less progress on the higher dimensional optimization problem in the limited number of function evaluations that are allowed before the solver terminates}.
\\
\\
The effect of the number of parameters drawn from the prior distribution, $N$, is shown in Figures \ref{figEnsembleRedRob} and \ref{figEnsembleIncrRob}. For the non-robust experiment, the mean of the prior distribution of the model parameters was used in the calculation of the Fisher information matrices in Equations (\ref{expectedFIMk}) and (\ref{observedFIM}), instead of a single random value from this distribution. The effect of reducing robustness is much less pronounced than the effect of reducing the control horizon. Only the convergence of the estimate of the damper constant is slower. Increasing the number of draws from the prior distribution from $100$ to $400$ also seems to have little added value past a certain point. In fact, it seems to perform slightly worse, we are unsure of the reason why.
\\
\\
We also compare our adaptive experimental design technique to the non-adaptive strategy from Equation (\ref{bayesian optimality}), in Figure \ref{figEnsembleNonAdp}. In the beginning of the experiment, the non-adaptive experiment performs equally well as our adaptive strategy. However, later on in the experiment, between time-steps $50$ and $100$, the estimation of the damper constant remains hovering slightly too high instead of continuing to converge to the true value. An additional shortcoming of the non-adaptive design is the large computational time required to generate this design. This is because the non-adaptive experiment requires predicting $100$ steps into the future. The optimization of this design was much slower than the adaptive designs with a short control window, due to the presence of large matrices in the expected FIM in Equation (\ref{expectedFIMk}).
\\
\\
Finally, the effect of a larger number of time-steps is shown in Figure \ref{figEnsembleIncrTime}. The figure shows that, also here, there are diminishing marginal returns for longer experiments. {\color{black}This figure contains strange spikes after $200$ time-steps. These spikes occur when the log-likelihood of all $\bm \theta_i$ overflows and becomes equal to minus infinity. The algorithm does not know which $\bm \theta_i$ to pick as the maximum likelihood estimate. In future research, we will consider adding an early stopping criterion for Algorithm \ref{algo}, when the log-likelihood of all $\bm \theta_i$ overflows.}
{\color{black}
\subsection{Two Compartment System}
In the second case study, we consider experimental design for a two compartment system. The discretized linear dynamics of this system are:
\begin{equation}
	\begin{aligned}
		\bm x_k &= \begin{bmatrix}
			1 - \Delta t(K_{1,0} + K_{1,2})  & \Delta tK_{2,1}\\
			\Delta tK_{1,2} & 1 - \Delta tK_{2,1}
		\end{bmatrix} \bm x_{k-1} +
		\begin{bmatrix} \Delta t \\ \frac{\Delta t^2}{2}\end{bmatrix} u_k + 
		\bm w_k,\\
		\bm w_k & \sim  \mathcal{N}\left(\begin{bmatrix}0\\0\end{bmatrix},
		\begin{bmatrix} q \Delta t & \frac{q\Delta t^2}{2} \\\frac{q\Delta t^2}{2}  &\frac{q\Delta t^3}{3} \end{bmatrix}\right),\\
		y_k &= \begin{bmatrix} \Delta tK_{1,0} & 0 \end{bmatrix} \bm x + v_k, \\
		v_k &\sim \mathcal{N}(0,0.0001).
	\end{aligned}
\end{equation}
In these equations, $K_{1,2}$, $K_{2,1}$ and $K_{1,0}$ are the unknown model parameters that determine the flows between the two compartments and the flow from the first compartment to the environment. Their true values all equal $0.2$. The prior distributions we use for them are independent normal distributions centered around $0.22$, with variances equal to $0.0016$. The time between measurements, $\Delta t$, is equal to $0.1$. The outflow of the first compartment to the environment is the measured output, with measurement noise on top of it. The initial distributions for the two compartments are independent normal distributions with means equal to $10$ and $1$, and variances equal to $0.01$ and $0.00001$, respectively. The controllable input at time-step k, $u_k$, is a flow towards the first compartment. This input is constrained between $0$ and $10$. There is also an unknown stochastic input $\bm w_k$ to the first compartment, represented by a discretization of Brownian motion with spectral density $q$ equal to $0.00625$. Brownian motion is a continuous time stochastic process whose increments are independent, stationary and normally distributed. The variance of the increments is determined by the spectral density. See \textcite{solin} for a more technical definition of Brownian motion and spectral density. The variance of the measurement noise is equal to $0.000625$.
\\
\\
This example shows the added value of working with arbitrarily parameterized state space models, instead of linear autoregressive models. Some parameters, such as $K_{1,0}$, occur multiple times in the system dynamics. This is contrary to autoregressive models, where the output is assumed to be a linear combination of previous measurements and inputs, and each parameter of this linear combination is allowed to vary freely in the parameter estimation.
\\
\\
Figure \ref{figOnline2} depicts the progression of the online maximum likelihood estimate as time goes on. The optimal experiment was generated with $N=1000$ draws from the prior distribution, and looks $e=3$ steps ahead. The parameters are estimated precisely after roughly $150$ time steps, while the random experiment does not correctly estimate the model parameters even after $200$ steps.
\begin{figure}[H]
	\begin{subfigure}[b]{0.90\textwidth}
		\centering
		\includegraphics[width=0.8\textwidth]{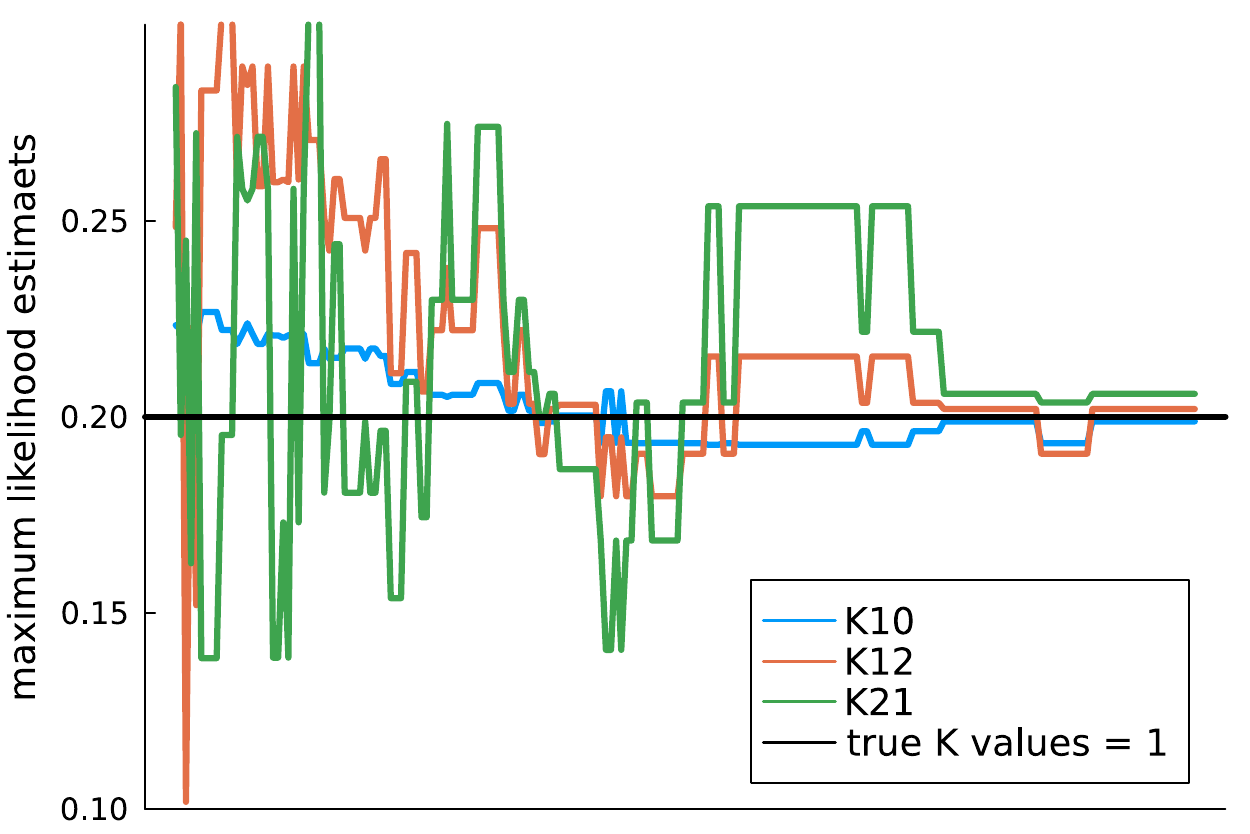}
		\caption{Input.}
	\end{subfigure}\\
	\begin{subfigure}[b]{0.90\textwidth}
		\centering
		\includegraphics[width=0.8\textwidth]{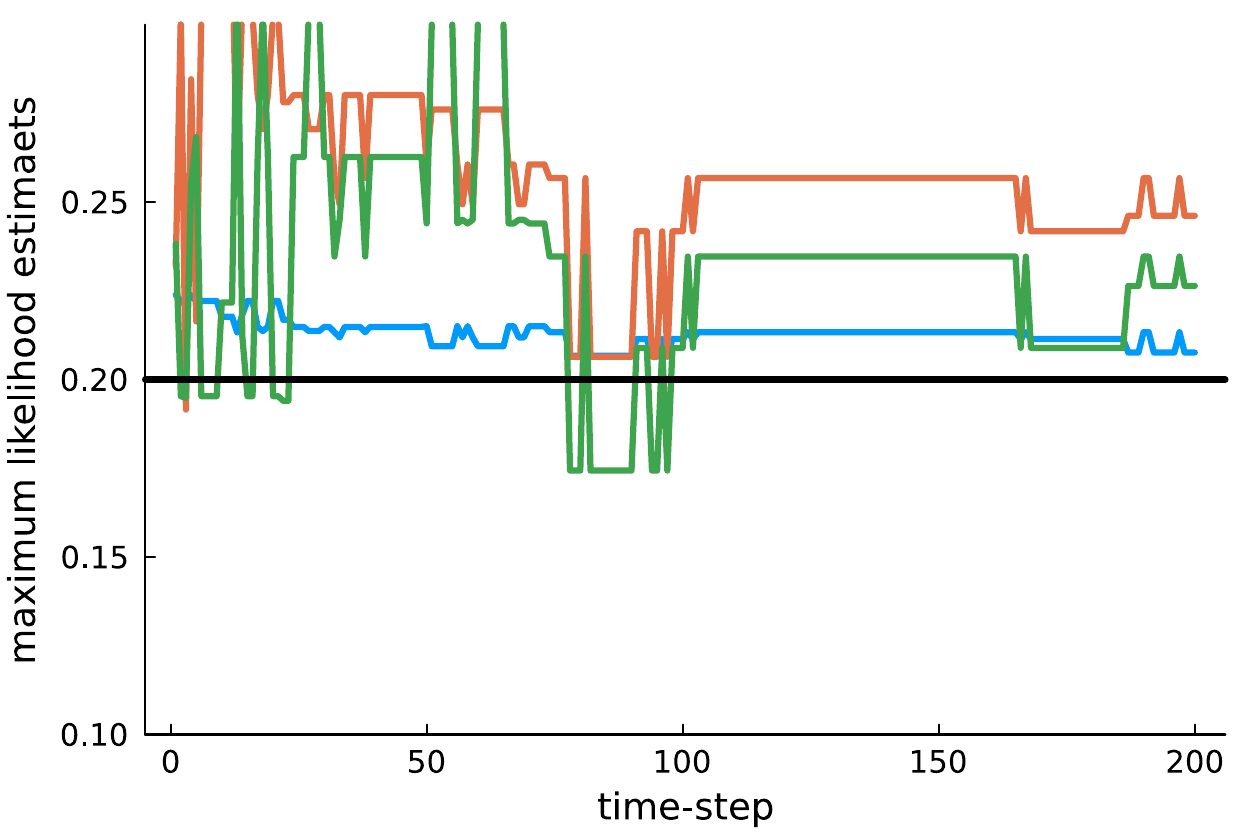}
		\caption{Output.}
	\end{subfigure}
	\caption{Online maximum likelihood estimates for the two compartment model.}
	\label{figOnline2}
\end{figure}
}
\section{Discussion and Conclusion}
In this paper, we presented a novel robust and adaptive experimental design method to estimate the model parameters of discrete-time linear state space models. We achieved this by quantifying the information content of an experiment using a combination of the expected and observed Fisher information matrix. In future research, we want to extend these results to non-linear dynamics. The Kalman filter must then be replaced with another Bayesian filter, such as the extended Kalman filter, sigma-point filter or particle filter.
\\
\\
{\color{black}We explicitly calculated the likelihood of the static parameters every time-step in Equation (\ref{likelihood factor}), but another method to estimate these parameters, is to append them to the dynamical system \parencite{sarkka}. The state and static parameters can then be estimated by a single non-linear filter. Often the extended Kalman filter is used for this. However, this filter forces a Gaussian approximation on the estimate of the static parameters. If we do not want to use a Gaussian approximation for the uncertainty in the static parameters, we could use a particle filter. But a complete Monte Carlo approach, like the particle filter, is wasteful, since it does not exploit the linearity present in the system dynamics, given the static parameters. The mixture Kalman filter does exploit this property \parencite{chen}. In the mixture Kalman filter some states are approximated by particles, and for each particle, a Kalman filter keeps track of the remaining states. This seems very similar to our Equation (\ref{Kalman}). Further investigation is needed to compare our method to the mixture Kalman filter.}
\\
\\
Continuous-time dynamics is another interesting direction for future research, as {\color{black}little} literature exists on experimental design for stochastic differential equation models, {\color{black} particularly when no analytical solution exist, and the model must be simulated by numerical techniques. For stochastic differential equation models which do have an analytical solution, some optimal design techniques have been developed by \textcite{anisimov,fedorov2}.}
\\
\\
To be able to optimize our experiments adaptively, we were forced to evaluate the likelihood of the model parameters at the same location (in the model parameters space) at every time-step. Most locations quickly become very unlikely, as seen in Figure \ref{figLikelihoodOpt}. It would be better if this sample could slightly move towards regions of higher probability at every time-step. \textcite{kantas} give an overview of methods that allow for such jittering of the location of the model parameters, but none of the methods discussed are completely online. Since these authors are only interested in online parameter estimation, {\color{black}and not both online estimation and experimental design, }this is a {\color{black}smaller} issue for them. But when considering adaptive experimental design, where an optimization over the input space has to be ran at every time-step, the jittering of the model parameters must be very efficient. Very recently \textcite{he}, published a promising method for online parameter estimation for linear dynamical systems based on the Kalman filter, that could be used for this purpose, and which we plan to incorporate in future research. {\color{black}Taking a higher quality initial sample of possible model parameters, is another way to remedy only being able to evaluate the likelihood at the same locations throughout the experiment. We took a Monte Carlo approach, randomly drawing $N$ values from the prior $p(\bm \theta)$, but it is worthwhile to investigate if taking a quasi-Monte Carlo approach like \textcite{teymur} would lead to better results.}
\printbibliography
\end{document}